\documentclass[%
superscriptaddress,
showpacs,
nofootinbib,
amsmath,amssymb,
aps,
prc,
twocolumn,
floatfix ]%
{revtex4-1}

\usepackage{comment}
\usepackage[dvips]{graphicx}
\usepackage{braket}
\usepackage{cases}
\usepackage{caption}
\usepackage[top=25truemm,bottom=20truemm,left=20truemm,right=20truemm]{geometry}

  \makeatletter
    
    \@addtoreset{equation}{section}
  \makeatother

\begin{document}

\renewcommand{\thefootnote}{\fnsymbol{footnote}}

\title{Comparative study of the requantization of the time-dependent mean
field for the dynamics of nuclear pairing}
\author{Fang Ni}
 \affiliation{Faculty of Pure and Applied Sciences,
              University of Tsukuba, Tsukuba 305-8571, Japan}
\author{Takashi Nakatsukasa}
 \affiliation{Center for Computational Sciences,
              University of Tsukuba, Tsukuba 305-8577, Japan}
 \affiliation{Faculty of Pure and Applied Sciences,
              University of Tsukuba, Tsukuba 305-8571, Japan}
 \affiliation{iTHES Research Group, RIKEN, Wako 351-0198, Japan}

\begin{abstract}
To describe quantal collective phenomena, it is useful
to requantize the time-dependent mean-field dynamics.
We study the time-dependent Hartree-Fock-Bogoliubov (TDHFB) theory
for the two-level pairing Hamiltonian, and compare results
of different quantization methods.
The one constructing microscopic wave functions,
using the TDHFB trajectories fulfilling the
Einstein-Brillouin-Keller quantization condition,
turns out to be the most accurate.
The method is based on the stationary-phase approximation
to the path integral.
We also examine the performance of the collective model
which assumes that the pairing gap parameter is the collective coordinate.
The applicability of the collective model is limited
for the nuclear pairing with a small number of
single-particle levels,
because the pairing gap parameter represents only a half of the
pairing collective space.
\end{abstract}

\maketitle

\section{Introduction}

Pairing correlation plays a decisive role in a number of nuclear phenomena,
which is especially important in open-shell nuclei.
Many evidences of the pairing correlation were observed in experiment,
including odd-even mass difference, moments of inertia of rotational bands,
and quasi-particle spectra in odd nuclei.
Even in closed-shell nuclei, the pairing dynamically plays an important
role in elementary modes of excitation, such as pairing vibrations
\cite{BM75, RS80, BB05}.
Properties of low-lying modes of excitation in even-even nuclei are
expected to be determined dominantly by interplay between
the pairing and the quadrupole correlations.
However, the true nature of the low-lying excitations is still unclear,
especially for excited $J^\pi=0^+$ states \cite{garrett,HW11}.
Understanding the pairing dynamics is a key ingredient for solving
mysteries of excited $0^+$ states.

The ground states in many of even-even nuclei are well
described by the Bardeen-Cooper-Schrieffer (BCS)
and the Hartree-Fock-Bogoliubov (HFB) theories \cite{RS80,BHR03}.
Its time-dependent version, 
the time-dependent HFB (TDHFB) theory \cite{BR86,NMMY16},
is a natural extension of the static HFB theory.
Thanks to increasing computational power,
realistic applications of the TDHFB calculations in real time
become available for studies of linear response properties
\cite{Eba10,SBMR11,Has12,SL13-2,ENI14}
and of various nuclear dynamics
\cite{SL13-1,SBBMR15,SSL15,WSMBF16,MSW17,HS17}.
The small-amplitude limit of the TDHFB is known to be 
the quasiparticle random phase approximation (QRPA).
The QRPA has been extensively utilized and successfully
describes properties of giant resonances.
Recently, the QRPA calculations with modern energy density functionals
for giant resonances in deformed nuclei have become available
\cite{Nes06,YG08,PG08,AKR09,Los10,TE10,YN11,YN13}.
In contrast, many of low-lying excited states
cannot be well reproduced by QRPA \cite{NMMY16}.
This may be due to their slowly moving large-amplitude nature.


In principle, the TDHFB dynamics can be applicable to large amplitude motion.
The problem is that it is not easy to determine quantum mechanical 
quantities, such as energy eigenvalues and transition matrix elements,
from the TDHFB trajectories.
In addition, the TDHFB lacks a part of quantum fluctuation
associated with low-energy
large amplitude collective motion,
which leads to difficulties in description of the quantum tunneling
processes, such as spontaneous fission and sub-barrier fusion reaction.
For this purpose,
since the TDHF(B) trajectory corresponds to a stationary-phase limit of the path
integral formulation,
the requantization of the mean-field dynamics was
proposed \cite{Neg82,L80,LNP80,KS80,K81,Rei80}.
It recovers quantum fluctuations missing in the mean-field
level, and possibly enables us to describe large-amplitude collective
tunneling phenomena.
The requantization of the TDHFB is particularly feasible
for integrable systems, because the system is described by separable
action-angle variables $(I_k, \phi_k)$, leading to
the Einstein-Brillouin-Keller (EBK) quantization condition.
However, for nonintegrable systems in general, it is difficult to find
suitable periodic orbits to quantize.
A possible solution to this difficulty is to find a decoupled collective
subspace spanned by
a single coordinate and its conjugate momentum \cite{NMMY16}.
Since the one-dimensional system is integrable, the quantization is
practicable.

Another somewhat phenomenological approach
to nuclear collective dynamics is the collective model.
In this approach, the collective Hamiltonian is constructed by choosing
collective coordinates intuitively.
The nuclear energy density functional model is often used as a tool for
a microscopic derivation of the collective Hamiltonian \cite{NMMY16}.
The most well-known and successful model is the Bohr model \cite{BM75},
which was introduced to describe low-energy
nuclear collective motion in quadrupole degrees of freedom with
the deformation parameters $(\beta, \gamma)$ and the Euler angles
$(\phi,\theta,\psi)$.
For the pairing motion, the collective coordinates are analogously
chosen as the pair deformation (gap) $\Delta$ and the gauge angle $\Phi$
\cite{BBPK70}.
Based on the pairing collective Hamiltonian, 
effects of the pair motion on the quadrupole vibrations 
have been discussed in literature \cite{delta1,delta3}.
However, it is not trivial whether the pair deformation is really a suitable
choice for the collective coordinate,
which should be investigated in a microscopic approach based on
the TDHFB dynamics.

Our final goal is to study the role of large-amplitude pairing dynamics
and to reveal nature of the mysterious excited $0^+$ states.
As the first step, toward this goal, we investigate accuracy and
applicability of the requantized TDHFB model for a two-level pairing model
with equal degeneracy $\Omega$ \cite{HF60},
especially on the calculation of two-particle-transfer matrix elements.
In Ref.~\cite{CDS84},
the two-particle-transfer matrix elements were evaluated as
Fourier components of the time-dependent mean values of the pair-creation operators,
which demonstrates nice agreement with the exact results at
large $\Omega$ values ($\Omega=40$).
However, in realistic values of $\Omega$, we will show that
the deviation is substantial.
The collective model of the pairing motion, which assumes the pair deformation
as the collective coordinate, has a similar tendency, namely, 
applicability limited to large $\Omega$ cases \cite{BBPK70}.
This deficiency is mainly due to the small collectivity in the pairing motion
in realistic situations.
In this paper, in order to improve the quantitative estimate of the matrix
elements, we construct microscopic wave functions
based on the EBK quantization for the integrable systems.
The wave functions are obtained
from the stationary-phase approximation for the path-integral form \cite{SM88}.
Its superiority to the other methods
becomes more evident for smaller values of $\Omega$.

The paper is organized as follows. 
In Sec. \ref{sec:TDHFB},
we derive a TDHFB classical Hamiltonian for the pairing model.
In Sec. \ref{sec:requantization}, 
the requantization of the TDHFB is performed using different methods,
based on the canonical and the EBK quantization.
In Sec. \ref{sec:results}, 
the numerical results for the two-level model are shown and compared with
exact results.
Properties of the pairing collective coordinate is also discussed.
We give a conclusion in Sec. \ref{sec:conclusion}.

\section{Classical form of TDHFB Hamiltonian}
\label{sec:TDHFB}

The Hamiltonian of the pairing model is given in terms of
single-particle energies $\epsilon_l$ and the pairing strength $g$ as
\begin{align}
	H &= \sum_l \epsilon_l n_l - g \sum_{l,l'} S_l^+ S_{l'}^- \nonumber \\
    &= \sum_l\epsilon_l(2S_l^0+\Omega_l) - g S^+ S^{-} ,
\end{align}
where we use the SU(2) quasi-spin operators,
$\boldsymbol{S}=\sum_l \boldsymbol{S}_l$, with
\begin{eqnarray}
        S_l^0 &=& \frac{1}{2}(\sum_ma_{lm}^{\dag}a_{lm}-\Omega_l) ,\\
        S_l^{+} &=& \sum_{m>0}a_{lm}^{\dag}a_{l\overline{m}}^{\dag} ,
\quad   S_l^{-} = S_l^{+\dag} .
\end{eqnarray}
Each single-particle energy $\epsilon_l$ possesses $(2\Omega_l)$-fold
degeneracy ($\Omega_l=j_l+1/2$)
and $\sum_{m>0}$ indicates the summation over $m=1/2,3/2,\cdots,$ and $\Omega_l-1/2$.
The occupation number of each level $l$ is given by
$
	n_l = \sum_m a^{\dag}_{lm}a_{lm} = 2S_l^0+\Omega_l ,
$.
The quasi-spin operators satisfy the commutation relations
\begin{equation}
  [S_l^0,S_{l'}^{\pm}] = \pm\delta_{ll'}S_{l}^{\pm},
	\quad [S_{l}^{+},S_{l'}^{-}] = 2\delta_{ll'}S_{l}^{0} .
\end{equation}
The magnitude of quasi-spin for each level is
$S_l=\frac{1}{2}(\Omega_l-\nu_l)$, where $\nu_l$ is the seniority
quantum number, namely the number of unpaired particle at each level $l$.
In the present study, we only consider seniority zero states with
$\nu=\sum_l \nu_l=0$.
The residual two-body interaction only consists of monopole pairing
interaction which couples two particles to zero angular momentum.
We obtain exact solutions either by solving Richardson equation
\cite{Richardson,Richardson2,Richardson3} or
by diagonalizing the Hamiltonian using the quasi-spin symmetry.

\subsection{Coherent-state representation of the TDHFB Hamiltonian}

The coherent state for the seniority $\nu=0$ states
($S_l=\Omega_l/2$) is constructed as
\begin{equation}
	\ket{Z(t)} = \prod_{l} \left(1+|Z_l(t)|^2\right)^{-\Omega_l/2}
	\exp [Z_l(t) S_l^{+}] \ket{0}
 \label{coherent}
\end{equation}
where $\ket{0}$ is the vacuum (zero particle) state,
$Z_l(t)$ are time-dependent complex variables which describe
motion of the system. 
In the SU(2) quasi-spin representation,
$\ket{0}=\prod_l \ket{S_l,-S_l}$.
The coherent state $\ket{Z(t)}$ is a superposition of
states with different particle numbers
without unpaired particles.
In the present pairing model,
the coherent state is the same as the time-dependent BCS wave function
with $Z_l(t)=v_l(t)/u_l(t)$,
where $(u_l(t),v_l(t))$ are the time-dependent BCS $u,v$ factors.

The TDHFB equation can be derived from the time-dependent variational
principle,
\begin{equation}
	\delta \mathcal{S} = 0,
	\quad
	\mathcal{S}\equiv
	\int \braket{\phi(t)|i\frac{\partial}{\partial t}-H|\phi(t)}dt,
  \label{TDHFB}
\end{equation}
where $\ket{\phi(t)}$ is the time-dependent generalized Slater determinant.
In the present case, we adopt the coherent state of Eq. (\ref{coherent}),
$\ket{\phi(t)}=\ket{Z(t)}$.
The action $\mathcal{S}$ is
\begin{align}
	\mathcal{S}  &= \int \mathcal{L}(t) dt \nonumber \\
	&= \int dt \left\{ \frac{i}{2} \sum_l \frac{\Omega_l}{1+|Z_l|^2}
  (Z_l^*\dot{Z_l}-Z_l\dot{Z_l^*}) - \braket{Z|H|Z} \right\} ,
\label{S}
\end{align}
and
\begin{align}
  \braket{Z|H|Z} =& \sum_l \epsilon_l \frac{2\Omega_l |Z_l|^2}{1+|Z_l|^2}
  -g \sum_l \frac{\Omega_l|Z_l|^2(\Omega_l+|Z_l|^2)}{(1+|Z_l|^2)^2} \nonumber \\
  &-g \sum_{l_1 \neq l_2} \frac{\Omega_{l_1}Z_{l_1}}{1+|Z_{l_1}|^2}\frac{\Omega_{l_2}Z_{l_2}^*}{1+|Z_{l_2}|^2}.
	\label{TDHFB_Hamiltonian_1}
\end{align}
We transform the complex variables $Z_l$ into real variables $(q_l,\chi_l)$
by $Z_l = \tan{\frac{\theta_l}{2}}e^{-i\chi_l}$ and $q_l=\cos{\theta_l}$ ($0\leq\theta\leq\pi$).
The Lagrangian $\mathcal{L}$ and the expectation value of Hamiltonian become
\begin{align}
\mathcal{L}(t) = \sum_l \frac{\Omega_l}{2}
	(1&-q_l)\dot{\chi_l} - \mathcal{H}(Z,Z^*) ,\\
	\mathcal{H}(Z,Z^*) \equiv \braket{Z|H|Z}& \nonumber \\
  = \sum_l \epsilon_l\Omega_l(1- q_l)& - \frac{g}{4}\sum_l \Omega_l [\Omega_l(1-q_l^2)+(1-q_l)^2] \nonumber \\
- \frac{g}{4}\sum_{l_1\neq l_2} \Omega_{l_1}\Omega_{l_2}&\sqrt{(1-q_{l_1}^2)(1-q_{l_2}^2)}e^{-i(\chi_{l_1}-\chi_{l_2})}   .
\label{TDHFB_Hamiltonian_2}
\end{align}
Here, $\chi_l$ represents a kind of gauge angle of each level,
and $q_l$ are related to the occupation probability, $q_l=|u_l|^2-|v_l|^2$.
If we choose $\chi_l$ as canonical coordinates, their conjugate momenta
are given by
$p_l\equiv \partial\mathcal{H}/\partial\dot{\chi}_l=\Omega_l(1-q_l)/2$.
Since the Hamiltonian (\ref{TDHFB_Hamiltonian_2}) depends only on
relative difference in the gauge angles,
the ``global'' gauge angles, $\Phi\propto\sum_l \chi_l$,
is a cyclic variable.

\subsection{Two-level case}

In a two-level system,
it is convenient to define global and relative gauge angles,
$\Phi$ and $\phi$, respectively.
\begin{align}
  \Phi &\equiv \frac{\chi_1 + \chi_2}{2}, \quad\quad
\phi\equiv \chi_2 - \chi_1,
\label{phi}
\end{align}
whose ranges are $0\leq \Phi \leq 2\pi$ and $-2\pi \leq \phi \leq 2\pi$.
Their conjugate momenta $(J,j)$ are given by
\begin{align}
	J &= \frac{\partial\mathcal{L}}{\partial\dot{\Phi}} =  \sum_{l=1}^2 \frac{\Omega_l}{2}(1-q_l), 
\nonumber \\
j &= \frac{\partial\mathcal{L}}{\partial\dot{\phi}} = \frac{\Omega_2(1-q_2) - \Omega_1(1-q_1)}{4}.
	\label{pi}
\end{align}
By calculating the occupation number $n_l$ in the level $l$,
the physical meaning of these conjugate momenta becomes obvious
\begin{equation}
   n_l = \braket{Z|n_l|Z} 
	= \Omega_l(1-q_l) .
\end{equation}
Therefore, $J$ corresponds to the total particle number
$N=\sum_l n_l$,
while $j$ corresponds to the difference of occupation number
between upper level and lower level 
\begin{align}
  J &= \frac{N}{2}, \hspace{5mm} j = \frac{n_{2}-n_{1}}{4} .
  \label{momentum}
\end{align}
The Hamiltonian in terms of these canonical variables
$(\phi,j;\Phi,J)$ is given by
\begin{align}
\mathcal{H}(\phi,j;J)
	=& \sum_{l=1,2} \Omega_l \epsilon_l(1- q_l) \nonumber \\
	&- \frac{g}{4}\sum_{l=1,2} \Omega_l [\Omega_l(1-q_l^2)+(1-q_l)^2] \nonumber \\
&- \frac{g}{2}\Omega_1\Omega_2
	\sqrt{(1-q_1^2)(1-q_2^2)} \cos\phi
\label{TDHFB_Hamiltonian_3}
\end{align}
with
\begin{equation}
	q_l=\frac{\Omega_l - J -2(-1)^l j}{\Omega_l} 
	\quad \mbox{ for } l=1,2.
	\label{q_l}
\end{equation}
Note that the Hamiltonian does not depend on the global gauge angle $\Phi$.
This leads to the particle number conservation, $dN/dt=0$.



\begin{figure*}[tp]
 \begin{center}
  \includegraphics[width=60mm,angle=-90]{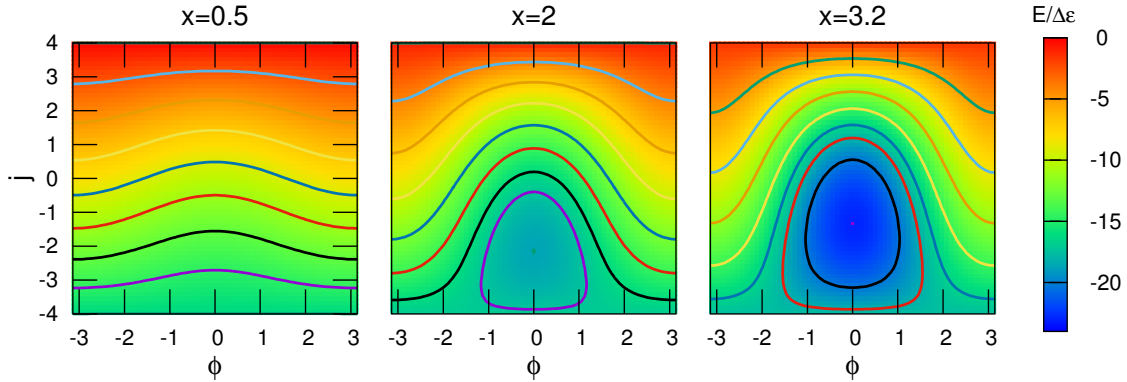}
 \end{center}
 \caption{Energy contour plot for $\Omega_1=\Omega_2=8$ and $N=16$.
	The lines indicate the TDHFB trajectories fulfilling
	the EBK quantization condition of Eq. (\ref{EBK}).
	}
 \label{fig:phase_space}
\end{figure*}

The TDHFB equation can be written in a form of
the classical equations of motion:
\begin{eqnarray}
	&&\frac{d\Phi}{dt} = \frac{\partial\mathcal{H}}{\partial J} ,\quad\quad
\frac{dJ}{dt} = -\frac{\partial\mathcal{H}}{\partial \Phi} , \\
	&&\frac{d\phi}{dt} = \frac{\partial\mathcal{H}}{\partial j} ,\quad\quad
\frac{dj}{dt} = -\frac{\partial\mathcal{H}}{\partial \phi} .
\label{TDHFB_equation}
\end{eqnarray}
Since the $J=N/2$ and the total energy $E$ are constants of motion,
the TDHFB trajectories with given $N$ and $E$ are determined
in the two-dimensional phase space $(\phi,j)$ with the condition
\begin{equation}
  \mathcal{H}(\phi(t),j(t);J=N/2) = E.
\end{equation}
Examples of the classical trajectories in the phase space
$(\phi,j)$ are shown in Fig.~\ref{fig:phase_space}.
The figure shows contour lines of energy for systems,
which correspond to the TDHFB trajectories,
with $\Omega_1=\Omega_2=8$ and $N=16$ with different values of $g$.
The transition from normal to superfluid phase takes place at
$g=\Delta\epsilon(2\Omega)^{-1}$ with $\Delta\epsilon=\epsilon_2-\epsilon_1$.
Using a dimensionless parameter
$
  x = 2g\Omega/\Delta\epsilon,
$
at $x<1$, the ground state is normal with fully occupied
lower level $n_1=N$ and empty upper level ($n_2=0$).
All the TDHFB trajectories represent rotational behavior with respect to
the relative angle $\phi$.
Here, the ``rotational behavior'' means that the motion spans the whole
region of the angle $\phi$,
while we use ``vibrational'' for the classical motion in a bound region of
$\phi$ ($-\pi<\phi<\pi$).
At $g=x=0$, the Hamiltonian becomes independent from $(\Phi,\phi)$, then,
the occupation numbers, $n_1$ and $n_2$, are constants of motion.
At $x>1$, the energy-minimum point and the closed trajectories appear around
$j=j_0$ ($-N/4 < j_0 < N/4$) and $\phi=0$,
which suggests the vibrational behavior for $\phi$.
At higher energies, the trajectories become open (rotational-like),
which suggests a phase transition from super to normal phases
as a function of excitation energy.

In the single-$j$ model, since the second term in the Hamiltonian
(\ref{TDHFB_Hamiltonian_3}) is absent,
the Hamiltonian is exactly quadratic with respect to $N$,
$\mathcal{H}(N) = \mathcal{H}_0+(N-N_0)^2/(2\cal{J})$.
The moment of inertia for the pair rotation is 
$\mathcal{J}=2/g\times\Omega/(\Omega-1)\approx 2/g$
at $\Omega\rightarrow\infty$.
In the multi-$j$ case, the Hamiltonian contains higher-order
terms in general.

\section{Requantization of TDHFB for the two-level pair model}
\label{sec:requantization}

In order to determine energy eigenstates,
we need to requantize the TDHFB trajectories.
Since the present two-level model is integrable (Sec.~\ref{sec:TDHFB}),
it is feasible to apply the stationary-phase approximation to
the path integral expression.
In addition, we also study the canonical quantization of the TDHFB
Hamiltonian, and the matrix elements extracted by the Fourier components
\cite{CDS84}.
We show results of these different approaches to the pairing model.

\subsection{Stationary phase approximation to the path integral}
\label{sec:SPA}

Starting an arbitrary state $\ket{\psi(0)}$ at time $t=0$,
the time-dependent full quantum state can be written in the path integral form
\begin{align}
\ket{\psi(t)}
	=& e^{-iHt} \ket{\psi(0)} \nonumber \\
	=& \int d\mu(Z'') \ket{Z''} \int d\mu(Z') \nonumber \\
	&\times \int_{Z(0)=Z'}^{Z(t)=Z''} \mathcal{D}\mu[Z(\tau)]
	e^{i\mathcal{S}[Z(\tau)]} \psi(Z') ,
\label{path_integral_expression}
\end{align}
where $\psi(Z)\equiv \braket{Z|\psi(0)}$ and
the invariant measure $d\mu(Z)$ is defined by 
the unity condition,
\begin{equation}
  \int d\mu(Z) \ket{Z}\bra{Z} = 1 .
  \label{unity}
\end{equation}
In Eq. (\ref{path_integral_expression}),
$\mathcal{S}[Z(\tau)]$ is the action (\ref{S}) along a given path $Z(\tau)$
with the initial coherent state $\ket{Z(0)}=\ket{Z'}$ and
the final state $\ket{Z(t)}=\ket{Z''}$,
then, the integration $\int \mathcal {D}\mu[Z(\tau)]$
is performed over all possible paths $\ket{Z(\tau)}$ between them.
Among all trajectories in the path integral,
the lowest stationary-phase approximation selects
the TDHFB (classical) trajectories\footnote{
In this formulation, the stationary-phase approximation agrees with
the TDHF(B) trajectories, while that to the auxiliary-field path
integral of Refs. \cite{Neg82,L80}
leads to the TDH(B) without the Fock potentials.
}.
\begin{equation}
	\ket{\psi(t)} \approx \int d\mu(Z') \ket{Z'_{\rm cl}(t)}
	e^{i\mathcal{S}_{\rm cl}(Z'_{\rm cl}(t),Z')} \psi(Z') ,
	\label{SPA}
\end{equation}
where the TDHFB trajectory starting from $\ket{Z'}$
ends at $\ket{Z'_{\rm cl}(t)}$ at time $t$.
The action $\mathcal{S}_{\rm cl}(Z_f,Z_i)$ is calculated along this
classical trajectory
connecting $Z_i=Z'_{\rm cl}(0)=Z'$ and $Z_f=Z'_{\rm cl}(t)$.
\begin{align}
	\mathcal{S}_{\rm cl}(Z'_{\rm cl}(t),Z') &\equiv \int_{0}^{t}
\braket{Z_{\rm cl}(t)|
	i\frac{\partial}{\partial t} - H
	|Z_{\rm cl}(t)} dt \nonumber \\
	&= \mathcal{T}[Z_{\rm cl}] - \mathcal{H}(Z',Z'^*) t 
	,
	\label{S_cl}
\end{align}
with
\begin{align}
\mathcal{T}[Z_{\rm cl}] &\equiv
\int_{0}^{t} \braket{Z_{\rm cl}(t)| i\frac{\partial}{\partial t}
	|Z_{\rm cl}(t)} dt \nonumber \\
	&= \int_{Z'}^{Z'_{\rm cl}(t)}
	\frac{i}{2} \sum_l \frac{\Omega_l}{1+|Z_l|^2}
  (Z_l^* dZ_l - Z_l dZ_l^*) .
\end{align}
In the last equation of Eq. (\ref{S_cl}),
we used the fact that the TDHFB trajectory conserves the energy,
$\mathcal{H}(Z_{\rm cl}(t),Z_{\rm cl}^*(t))= \mathcal{H}(Z',Z'^*)$.

The energy eigenstates correspond to stationary states,
$\braket{Z|\psi(t)} \propto \braket{Z|\psi(0)}=\psi(Z)$,
which can be constructed by superposing the coherent states along
a periodic TDHFB trajectory $Z_{\rm cl}^{(k)}$ as \cite{KS80,K81,SM88}
\begin{equation}
	\ket{\psi_k} = \oint d\mu(Z_{\rm cl}^{(k)}) \ket{Z_{\rm cl}^{(k)}}
	e^{i\mathcal{T}[Z_{\rm cl}^{(k)}]} .
\end{equation}
The single-valuedness of the wave function leads to
the quantization condition ($k$: integer)
\begin{align}
	\mathcal{T}_\circ[Z_{\rm cl}^{(k)}]=&
	\oint \frac{i}{2} \sum_l \frac{\Omega_l}{1+|Z_l^{(k)}|^2} \nonumber \\
	&\times \left(Z_l^{(k)*} dZ_l^{(k)} - Z_l^{(k)} dZ_l^{(k)*}\right) \nonumber \\
	=& 2k\pi .
	\label{EBK}
\end{align}
The state evolves in time as $\ket{\psi_k(t)}=\ket{\psi_k}e^{-iE_k t}$,
with the energy of the $k$-th periodic trajectory,
$E_k=\mathcal{H}(Z_{\rm cl}^{(k)},Z_{\rm cl}^{(k)*})$.

Finding TDHFB trajectories satisfying the quantization condition
(\ref{EBK}) is an extremely difficult task in general.
It is better founded and more practical
if the classical system is completely integrable.
In integrable systems, $M$ complex variables $Z(t)$ can be transformed into
the action-angle variables;
\begin{align}
	Z(t) &= \{Z_l(t);l=1,\cdots,M \} \nonumber \\
	&\rightarrow
	\{E; v_1,\cdots,v_{M-1}; \theta_1(t),\cdots,\theta_M(t)\} ,
\end{align}
where the variables $E$ and $v$ define an invariant torus,
while $\theta(t)$ parameterize the coordinates on the torus.
The integration path of Eq. (\ref{EBK}) is now taken as topologically
independent closed path on the torus,
namely the EBK quantization condition.
There are $M$ independent closed paths and $M$ quantum numbers,
$k=(k_1,\cdots,k_M)$, to specify the stationary energy eigenstate.
These are associated with $M$ invariant variables,
$\{E_k;v_1^{(k)},\cdots,v_{M-1}^{(k)}\}$.
Using the invariant measure
\begin{align}
  d\mu(Z) = \rho(E,v,\theta) dEdv_1\cdots dv_{M-1}d\theta_1\cdots d\theta_M,
\end{align}
the $k$-th semiclassical wave function can be calculated as
\begin{align}
  \ket{\psi_k} \propto \oint d\theta_1\cdots \oint d\theta_M &\ 
	\rho(E_k,v^{(k)},\theta)
	\ket{E_k,v^{(k)},\theta} \nonumber \\
	&\times e^{i\mathcal{T}[E_k,v^{(k)},\theta]} .
  \label{semi_wave_func}
\end{align}
Here, we omit the integration with respect to the invariant variables,
$E$ and $v$.


We apply semiclassical approach to the two-level pairing model.
The invariant measure in SU(2)$\otimes$SU(2) is 
\begin{align}
d\mu(Z) &= \prod_l \frac{\Omega_l+1}{\pi}(1+|Z_l|^2)^{-2}d{\rm Re}Z_ld{\rm Im}Z_l \\
  &= \prod_l \frac{\Omega_l+1}{4\pi}d\cos{\theta_l}d\chi_l \\
	&= \left(\prod_{l=1,2} \frac{1+\Omega_l^{-1}}{2\pi}\right)
             d\Phi dJ d\phi dj .
\end{align}
In the last equation, we transform the canonical coordinates
by Eqs. (\ref{phi}) and (\ref{pi}).
Since the particle number $J=N/2$ and the total energy $E$ are invariant,
the two-level pairing model is integrable.
Thus, we can construct the semiclassical wave function using
Eq. (\ref{semi_wave_func}).
The action integral is given by
\begin{align}
	\mathcal{T}_k(\Phi,\phi;J)
	&= J\Phi + \int^\phi_{-\pi} j' d\phi' 
	= \frac{N}{2}\Phi + \int^t_0   
	j(t') \frac{d\phi}{dt'} dt' \nonumber \\
	&\equiv \mathcal{T}_{N,E_k}(\Phi,t) ,
\end{align}
where the integration $\int j d\phi$ is performed on the $k$-th
closed trajectory of Eq. (\ref{EBK}), and
the variables $(\phi,j)$ are transformed into $(t,E)$.
The semiclassical wave function fulfilling the EBK quantization condition
becomes
\begin{align}
	\ket{\psi_k^N} &\propto \oint d\Phi \oint dt
	e^{i\mathcal{T}_{N,E_k}(\Phi, t)}
	\ket{\Phi,t}_{N,E_k} \label{semi_wave_func0} \\
  &\propto \sum_{m=0}^{J}C_m^{(E_k,J)}\ket{S_1,-S_1+m;S_2,-S_2+(J-m)} ,
	\label{semi_wave_func1}
\end{align}
with $S_l=\Omega_l/2$, $J=N/2$, and the coefficients
\begin{align}
  C_m^{(E_k,J)} =& \binom{J}{m}\int_0^T dt \nonumber \\
 &\times \exp{\left( i\int^t j(t') \dot{\phi}(t') dt'-i(J/2-m)\phi(t) \right)} \nonumber \\ 
 &\times A(q_1,S_1,m)A(q_2,S_2,J-m) ,\\
  A(q,S,m) &= \left(\frac{1-q}{2}\right)^{m/2}\left(\frac{1+q}{2}\right)^{S-m/2}\sqrt{\frac{(2S)!m!}{(2S-m)!}}, \hspace{5mm} \nonumber
\end{align}
where $T$ is the period of the closed trajectory.
The TDHFB-requantized wave functions (\ref{semi_wave_func1}) are
eigenstates of the total particle number.
This is due to the integration over the global gauge angle $\Phi$,
which makes the particle number projection not only for the ground state
but also for excited states.
See Appendix for detailed derivation of Eq.~(\ref{semi_wave_func1}). 

Since we obtain the microscopic wave function of every eigenstate,
the expectation values and the transition matrix elements for any operator
can be calculated in a straightforward manner.
In Sec.~\ref{sec:results}, we show those of the pair-addition operator
$S^+$ which characterize properties of the pair condensates.

%

Before ending this section, let us note the periodic conditions
of the coordinates and the quantization condition.
Since the two original variables, $(\chi_1,\chi_2)$,
are independent periodic variables of the period $2\pi$,
in addition to the trivial periodicity of $2\pi$ for $\Phi$ and of $4\pi$
for $\phi$,
we have periodic conditions for
$(\Phi,\phi)\rightarrow (\Phi\pm \pi,\phi\pm 2\pi)$ and
$(\Phi,\phi)\rightarrow (\Phi\pm \pi,\phi\mp 2\pi)$.
The former (latter) corresponds to $\chi_2\pm 2\pi$ ($\chi_1\pm 2\pi$)
with $\chi_1$ ($\chi_2$) being fixed.
For open TDHFB trajectory (e.g., Fig.~\ref{fig:phase_space}),
the quantization condition becomes
\begin{align}
 \mathcal{T}_{N,E_k}(\Phi\pm\pi,\phi\pm 2\pi;J) &= \mathcal{T}_{N,E_k}(\Phi,\phi;J) + 2m\pi \nonumber \\
	 \Leftrightarrow \quad\quad \pm \frac{N}{2} \pi + \int_{-\pi}^\pi j d\phi
	&= 2m\pi .
\end{align}
This leads to the following:
\begin{equation}
	\int_{-\pi}^\pi j d\phi = 
	\begin{cases}
		2k\pi & \mbox{ for } N=4n ,\\
		(2k+1)\pi & \mbox{ for } N=4n+2 ,
	\end{cases}
	\label{EBK_2}
\end{equation}
where $m$, $k$ and $n>0$ are integer numbers.

\subsection{Canonical quantization}
\label{sec:canonical}

The most common approach to the quantization of the nuclear collective model
is the canonical quantization \cite{BM75}.
In the pairing collective model, the canonical quantization
was adopted in previous studies \cite{BBPK70,delta1,delta3}.
Assuming magnitude and phase of the pairing gap as collective coordinates,
a collective Hamiltonian was constructed in the second order in momenta.
Then, the Hamiltonian was quantized by the canonical quantization
with Pauli's prescription.
In this section, we apply similar quantization method to
the TDHFB Hamiltonian (\ref{TDHFB_Hamiltonian_3}).
The main difference is that the collective canonical variables are
not assumed in the present case, but are obtained from the TDHFB dynamics
itself.\par



It is not straightforward to apply Pauli's prescription
to the present case, 
because the TDHFB Hamiltonian (\ref{TDHFB_Hamiltonian_3})
is not limited to the second order in momenta.
In the present study, 
we adopt a simple symmetrized ordering, as
\begin{align}
	H(\hat{\phi},\hat{j},\hat{J})
	=& \sum_{l=1,2} \Omega_l \epsilon_l (1-q_l) \nonumber \\
	&- \frac{g}{4} \sum_{l=1,2} \Omega_l (\Omega_l(1-q_l^2)+(1-q_l)^2)
	\nonumber \\
	&- \frac{g}{4}\Omega_{1}\Omega_{2}
	\left\{ \sqrt{(1-q_{1}^2)(1-q_{2}^2)}\cos{\hat{\phi}} \right. \nonumber \\
	&\left. + \cos{\hat{\phi}}\sqrt{(1-q_{1}^2)(1-q_{2}^2)} \right\} .
\label{canonical_quantized_H}
\end{align}
As in Eq. (\ref{q_l}), $q_l$ contain $J$ and $j$ which
are replaced by
\begin{equation}
	\hat{J} = -i\frac{\partial}{\partial\Phi},\quad\quad
	\hat{j} = -i\frac{\partial}{\partial\phi} .
\end{equation}
Since $\Phi$ is a cyclic variable, 
we write the collective wave function $\Psi(\Phi,\phi)$ as eigenstates
of the particle number $N$ in a separable form
\begin{equation}
  \Psi_k^{(N)}(\Phi,\phi) = 
	\frac{1}{\sqrt{2\pi}}e^{i\frac{N}{2}\Phi}\psi_k^{(N)}(\phi) .
\end{equation}
Then, the problem is reduced to the one-dimensional Schr\"{o}dinger
equation for the motion in the relative angle $\phi$.
the Schr\"{o}dinger equation
\begin{equation}
	H\left( \phi,-i\frac{d}{d\phi};\frac{N}{2} \right)
	\psi_k^{(N)}(\phi) = E_k^{(N)}\psi_k^{(N)}(\phi),
	\label{Schroedinger_eq}
\end{equation}

The wave function should have a periodic property with respect to
the variable $\phi$;
$\psi_k(\phi)=\psi_k(\phi+4\pi)$.
For the adopted simple ordering of Eq. (\ref{canonical_quantized_H}),
it is convenient to use the eigenstates of $\hat{j}$ as the basis
to diagonalize the Hamiltonian.
They are
\begin{equation}
	\chi_j(\phi) = \frac{1}{\sqrt{4\pi}} e^{i\phi j} ,
	\quad\quad\mbox{with $j$: integer or half integer} .
\end{equation}
Since the Hamiltonian (\ref{canonical_quantized_H}) contains only
terms linearly proportional to $e^{\pm i\phi}$,
the basis states $\chi_j$ with half-integer difference in $j$
are not coupled with each other.
Thus, the eigenstates of Eq. (\ref{Schroedinger_eq}) can be expanded as
\begin{equation}
	\psi_k^{(N)}(\phi) = 
	\sum_{j=j_{\rm min},j_{\rm min}+1,\cdots}^{j_{\rm max}}
	c_{k,j}^{(N)} \chi_j^{(N)}(\phi) .
	\label{CQ_eigenstate}
\end{equation}
According to the relation $j=(n_2-n_1)/4=(N-2n_1)/4$ in Eq. (\ref{momentum}),
we adopt the (half-)integer values of $j$ for $N=4n$ ($N=4n+2$)
with integer $n$.
This is consistent with the quantization condition (\ref{EBK_2}).
The coupling term with different $j$
in Eq. (\ref{canonical_quantized_H}) vanishes
for $n_l=0$ and $n_l=2\Omega_l$,
which restricts values of $j$ in a finite range of
$j_{\rm min}\leq j \leq j_{\rm max}$.

In order to estimate the two-particle transfer matrix elements,
we construct the corresponding operators as follows.
The classical form of matrix elements are obtained as
\begin{align}
  S^+&(\Phi,J;\phi,j) = \braket{Z|\hat{S}^+|Z} \nonumber \\
 &= \frac{1}{2}
\left( \Omega_1\sqrt{1-q_1^2}e^{-i\phi/2}+\Omega_2\sqrt{1-q_2^2}e^{i\phi/2} \right) e^{i\Phi}
	\label{Sp_mean_value} \\
  S^-&(\Phi,J;\phi,j) = \braket{Z|\hat{S}^-|Z} \nonumber \\
 &= \frac{1}{2}
\left( \Omega_1\sqrt{1-q_1^2}e^{i\phi/2}+\Omega_2\sqrt{1-q_2^2}e^{-i\phi/2} \right) e^{-i\Phi} .
	\label{Sm_mean_value}
\end{align}
Again, we adopt a simple symmetrized ordering for the quantization:
\begin{widetext}
\begin{align}
  S^\pm (\hat{\Phi},\hat{J};\hat{\phi},\hat{j})
	= \frac{1}{4}\left( \Omega_1\sqrt{1-q_1^2}e^{\mp i\hat{\phi}/2}
+\Omega_2\sqrt{1-q_2^2}e^{\pm i\hat{\phi}/2} \right) e^{\pm i\hat{\Phi}} 
+ \frac{1}{4}e^{\pm i\hat{\Phi}}\left( 
e^{\mp i\hat{\phi}/2}\Omega_1\sqrt{1-q_1^2}
+e^{\pm i\hat{\phi}/2}\Omega_2\sqrt{1-q_2^2} \right) .
\end{align}
\end{widetext}
The exponential factors $e^{\pm i\Phi}$ change
the total particle number $N\rightarrow N\pm 2$,
while $e^{\pm i\phi/2}$ change the relative numbers,
$n_2-n_1 \rightarrow n_2-n_1 \pm 2$.
Using these operators,
the pair-addition transition strengths are calculated as
\begin{widetext}
\begin{align}
B(P_{\rm ad};k\rightarrow k') &=
|\braket{N',k'|S^+(\hat{\Phi},\hat{J};\hat{\phi},\hat{j})|N,k}|^2
	\nonumber \\
&= \left| \frac{1}{2\pi} \int_{0}^{2\pi}d\Phi \int_{-2\pi}^{2\pi}d\phi 
	\psi^{(N')*}_{k'}(\phi) e^{-i\frac{N'}{2}\Phi}
	S^+(\hat{\Phi},\hat{J};\hat{\phi},\hat{j})
	\psi^{(N)}_{k}(\phi)e^{i\frac{N}{2}\Phi} \right|^2,
\end{align}
\end{widetext}
which automatically vanishes for $N'\neq N+2$.

\subsection{Fourier decomposition of time-dependent matrix elements}
\label{sec:Fourier}

The requantization and calculation of the matrix elements can be also
performed using the time-dependent solutions of the TDHFB.
It was proposed and applied to the two-level pairing model \cite{CDS84},
which we recapitulate in this section.

The TDHFB provides a time-dependent solution $Z(t)$ starting from
a given initial state $Z(0)$.
The energy eigenvalues and the corresponding closed trajectories
are determined from the EBK quantization condition (\ref{EBK}).
The pair transfer matrix elements are evaluated as the Fourier components
of the time-dependent mean values $S^\pm(t)=S^\pm(Z(t))$,
Eqs. (\ref{Sp_mean_value}) and (\ref{Sm_mean_value}).
Since the global gauge angle $\Phi$ is a cyclic variable,
the motion in the relative gauge angle $\phi$ is independent from $\Phi$.
Thus, we calculate the time evolution of $\phi(t)$, and
find the period of the $k$-th closed trajectory $T$
satisfying Eq. (\ref{EBK}).
Then, the Fourier component,
\begin{equation}
	\tilde{S}^\pm(E_k; \omega) = \frac{1}{T}\int_0^T dt
	e^{i\omega t} S^\pm(t) ,
\label{Fourier_decomposition}
\end{equation}
corresponds to the pair transfer matrix element from the state $k$ to $k'$
when $\omega=2\pi (k'-k)/T$.
The pair-addition transition strengths are calculated as
\begin{equation}
	B(P_{\rm ad};k\rightarrow k') 
	= \left|\tilde{S}^+\left(E_k;\frac{2\pi}{T}\Delta k \right) \right|^2 ,
\end{equation}
with $\Delta k=k'-k$.
In this approach, the transition between the ground states of
neighboring nuclei ($N\rightarrow N+2$) corresponds to
the stationary component ($k=0$ and $\Delta k=0$),
namely the expectation value in the BCS approximation.

The derivation  of Eq. (\ref{Fourier_decomposition})
is based on the wave packet in the
classical limit \cite{LL65}.
The TDHFB state is assumed to be a superposition of eigenstates $\ket{\phi_k^N}$
in a narrow range of energy $E_{k_0}-\Delta E<E_k<E_{k_0}+\Delta E$,
\begin{equation}
	\ket{Z(t)} = \sum_N \sum_k c_k^N \ket{\phi_k^N} e^{-iE_k t} ,
	\label{superposition}
\end{equation}
where the eigenenergies are evenly spaced and
the coefficients $c_k^N$ slowly vary with respect to $k$ and $N$.
The expectation value of $S^\pm$ is
\begin{equation}
	S^\pm(t) = \sum_{N} \sum_{k,k'} c_{k'}^{N+2*} c_k^N 
	\braket{\phi_{k'}^{N+2}|S^\pm|\phi_k^N} e^{i(E_{k'}-E_k)t} .
\end{equation}
The matrix element $\braket{\phi_{k'}^{N+2}|S^\pm|\phi_k^N}$
quickly disappears as $|k'-k|$ increases, while it stays almost constant
for the small change of $k$ and $N$ with $|k'-k|$ being fixed.
Thus, we may approximate $c_{k'}^{N+2}\approx c_k^N$,
$E_{k'}-E_k\approx \omega_0 \Delta k$, and
that $\braket{\phi_{k'}^{N+2}|S^\pm|\phi_k^N}\approx
\braket{\phi_{k_0+\Delta k}^{N+2}|S^\pm|\phi_{k_0}^N}$ 
\begin{align}
	S^\pm(t) &\approx \sum_{N} \sum_{k} \left|c_k^N\right|^2
	\sum_{\Delta k}
	\braket{\phi_{k_0+\Delta k}^{N+2}|S^\pm|\phi_{k_0}^N}
	e^{i\omega_0 \Delta k t}  \nonumber \\
	&= \sum_{\Delta k}
	\braket{\phi_{k_0+\Delta k}^{N+2}|S^\pm|\phi_{k_0}^N}
	e^{i\omega_0 \Delta k t}  ,
\end{align}
where $k_0$ is a representative index value of the superposition in 
Eq. (\ref{superposition}).
From this classical wave packet approximation,
we obtain Eq. (\ref{Fourier_decomposition}).
It is not trivial to justify the approximation for small values of $\Omega$
and for transitions around the ground state.

\section{Results}
\label{sec:results}

In this section, we study the seniority-zero states ($\nu_1=\nu_2=0$) in
the two-level system with equal degeneracy, $\Omega_1=\Omega_2=\Omega$.
Since all the properties are scaled with the ratio, $g/\Delta\epsilon$,
where $\Delta\epsilon$ is the level spacing
$\Delta\epsilon=\epsilon_2-\epsilon_1$,
we define a dimensionless parameter to control the strength of
the pairing correlation
\begin{align}
  x = 2\Omega\frac{g}{\Delta\epsilon} .
\label{x}
\end{align}
For sub-shell-closed systems with $N=2\Omega$ system,
the transition from normal ($x<1$) to superfluid ($x>1$) states
takes place at $x=1$.\footnote{
Strictly speaking, the phase transition takes place at $x=2\Omega/(2\Omega-1)$. }

We apply the requantization methods in Sec.~\ref{sec:requantization}.
In the following, the stationary-phase approximation to the path
integral in Sec.~\ref{sec:SPA} is denoted as ``SPA'',
the Fourier decomposition method (Sec.~\ref{sec:Fourier}) as ``FD'',
and the canonical quantization with periodic boundary condition 
(Sec.~\ref{sec:canonical}) as ``CQ''.
Note that the SPA and the FD produce the same eigenenergies which
are based on the EBK quantization rule.

\subsection{Large-$\Omega$ cases}

\begin{figure}[t]
 \begin{center}
(a)	 \includegraphics[height=0.44\textwidth,angle=-90]{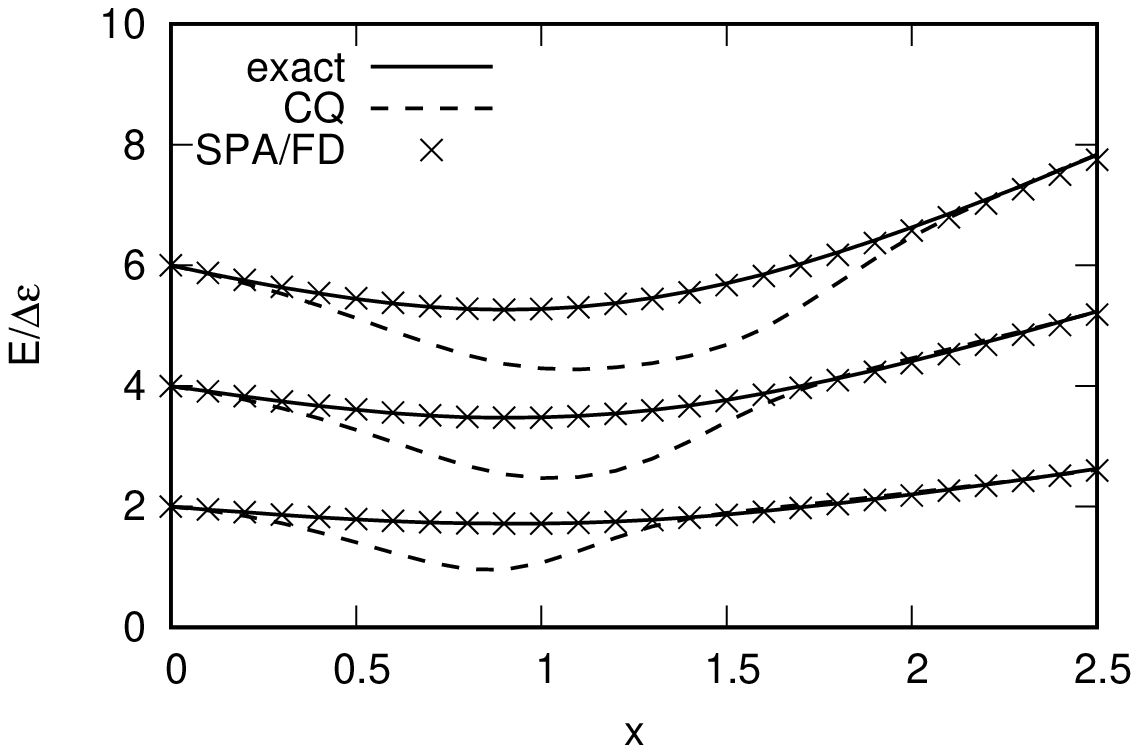}
\\
(b)	 \includegraphics[height=0.44\textwidth,angle=-90]{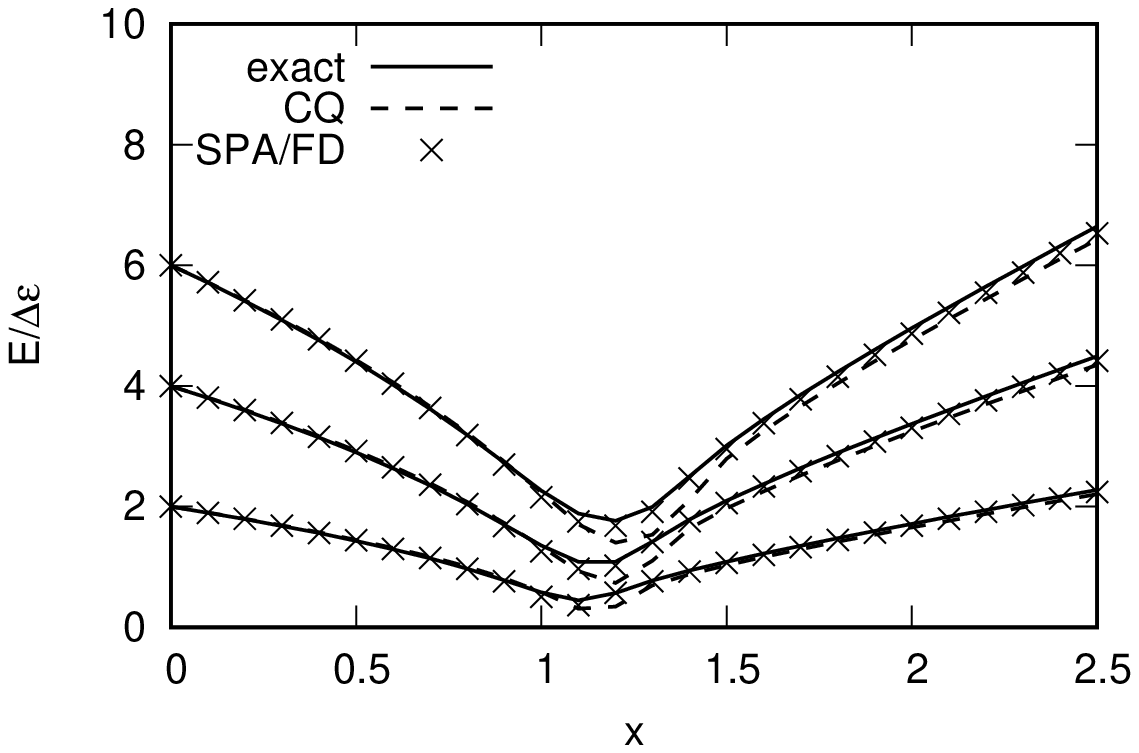}
 \end{center}
 \caption{Excitation energies of $\ket{0_2^+}$, $\ket{0_3^+}$ and
$\ket{0_4^+}$ for $\Omega=50$ systems with (a) $N=50$ and (b) $N=100$
	as functions of the dimensionless parameter $x$ of Eq. (\ref{x}).
}
 \label{fig:N50energy}
\end{figure}

In the limit of $\Omega\rightarrow\infty$,
we expect that the classical approximation becomes exact.
Here, we adopt $\Omega=50$ with $N=100$ (closed-shell configuration)
and $N=50$ (mid-shell configuration).

Calculated excitation energies are shown in Fig.~\ref{fig:N50energy}.
The results of SPA/FD and CQ are compared with the exact values.
At the weak pairing limit of $x\rightarrow 0$,
the excitation energies are multiples of $2\Delta\epsilon$, 
which correspond to pure $2n$-particle-$2n$-hole excitations.
Both the weak and the strong pairing limits
are nicely reproduced by all the calculations,
while the CQ method produces excitation energies slightly lower than the
exact values in an intermediate region around $x=1$.
It is somewhat surprising to see that the deviation is larger for
the case of the mid-shell configuration ($N=50$) than the closed shell
($N=100$).


The deviation in the CQ method is mainly due to the zero-point energy
in the ground state.
Since we solve the collective Schr\"odinger equation (\ref{Schroedinger_eq})
with the quantized Hamiltonian of Eq. (\ref{canonical_quantized_H}),
the zero-point energy $\Delta E>0$ is inevitable in the CQ method.
The $\Delta E$ is associated with the degree of localization of
the wave function.
Thus, the magnitude of $\Delta E$ for ``bound'' states are different
from that for ``unbound'' states.
See Fig.~\ref{fig:phase_space}.
In the strong pairing limit, the potential minimum is deep enough to bound
both ground and excited states.
Conversely, all the states are unbound in the weak limit.
In both limits, $\Delta E$ for ground and excited states are similar,
and they are canceled for the excitation energy.
However, near $x=1$, the ground state is bound,
while the excited states are unbound.
In this case, $\Delta E$ is larger in the ground state than in the
excited states, which makes the excitation energy smaller.
This also explains the difference between the mid-shell and closed-shell
configurations.
In the closed shell, all the states are unbound for $x<1$,
while, in the mid-shell,
there is a region in $x<1$ where the ground state is bound but
the excited state is unbound.

\begin{figure*}[t]
 \begin{minipage}{1\hsize}
 \begin{center}
  \includegraphics[height=0.45\textwidth,angle=-90]{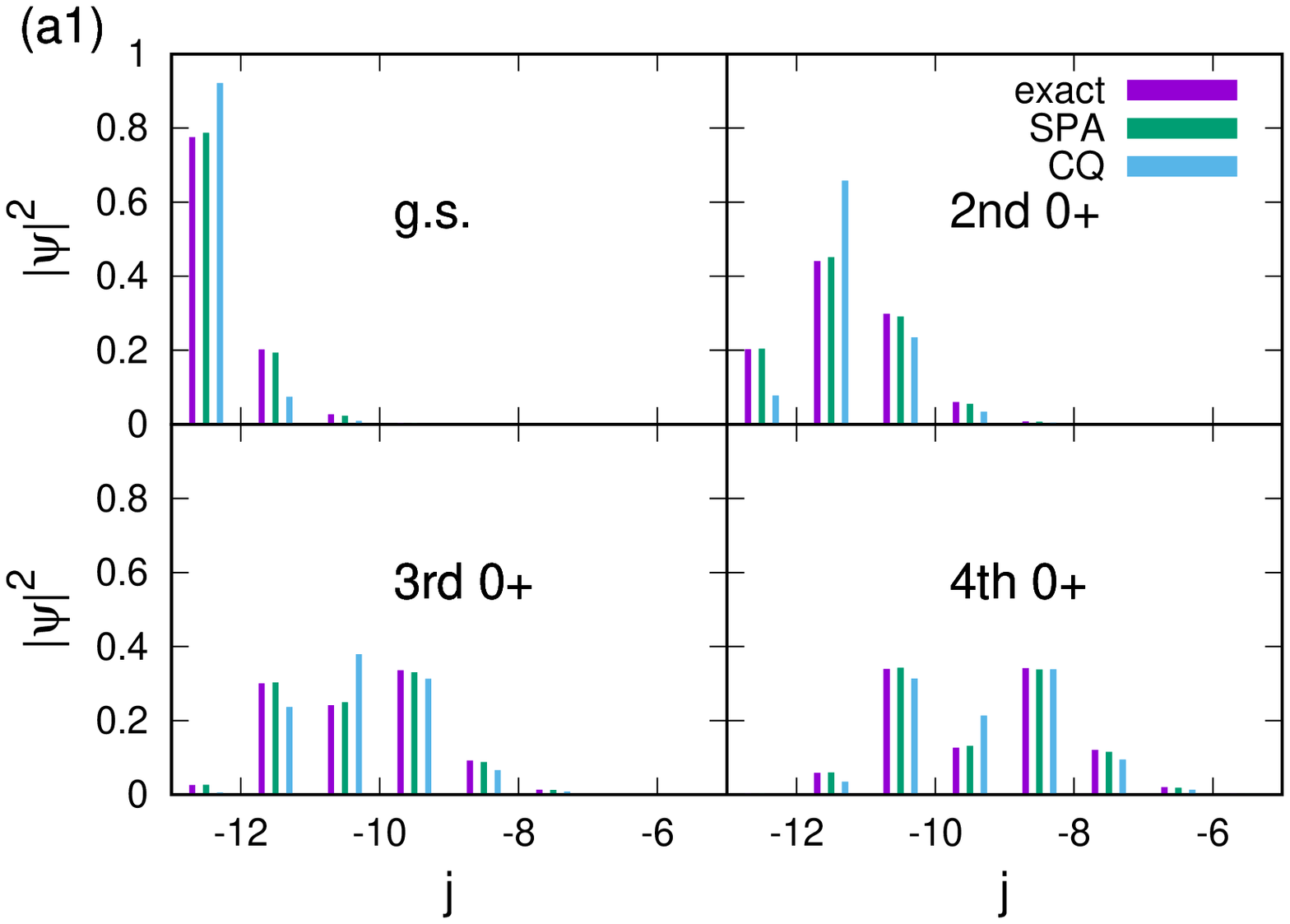}
  \includegraphics[height=0.45\textwidth,angle=-90]{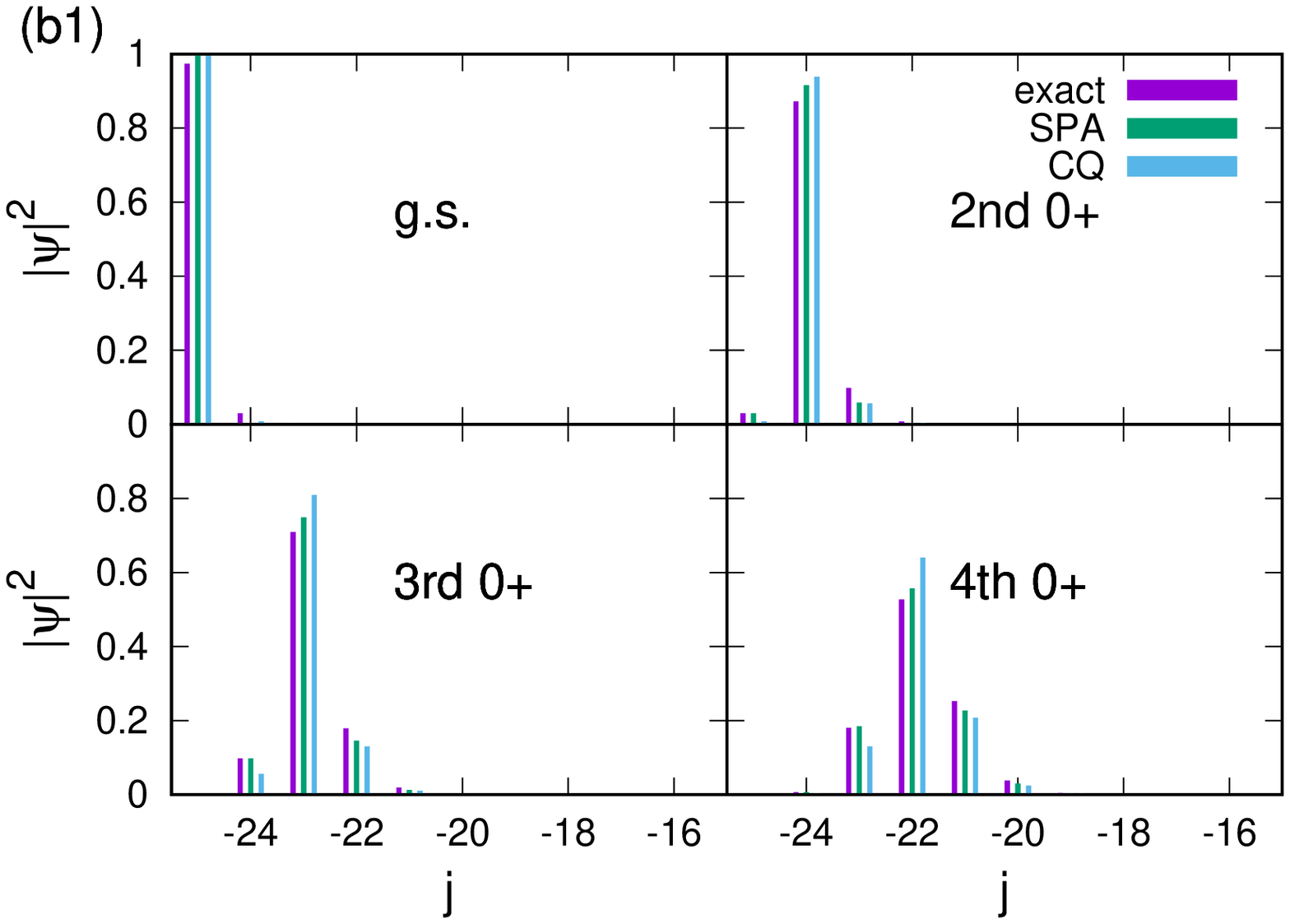}
  \includegraphics[height=0.45\textwidth,angle=-90]{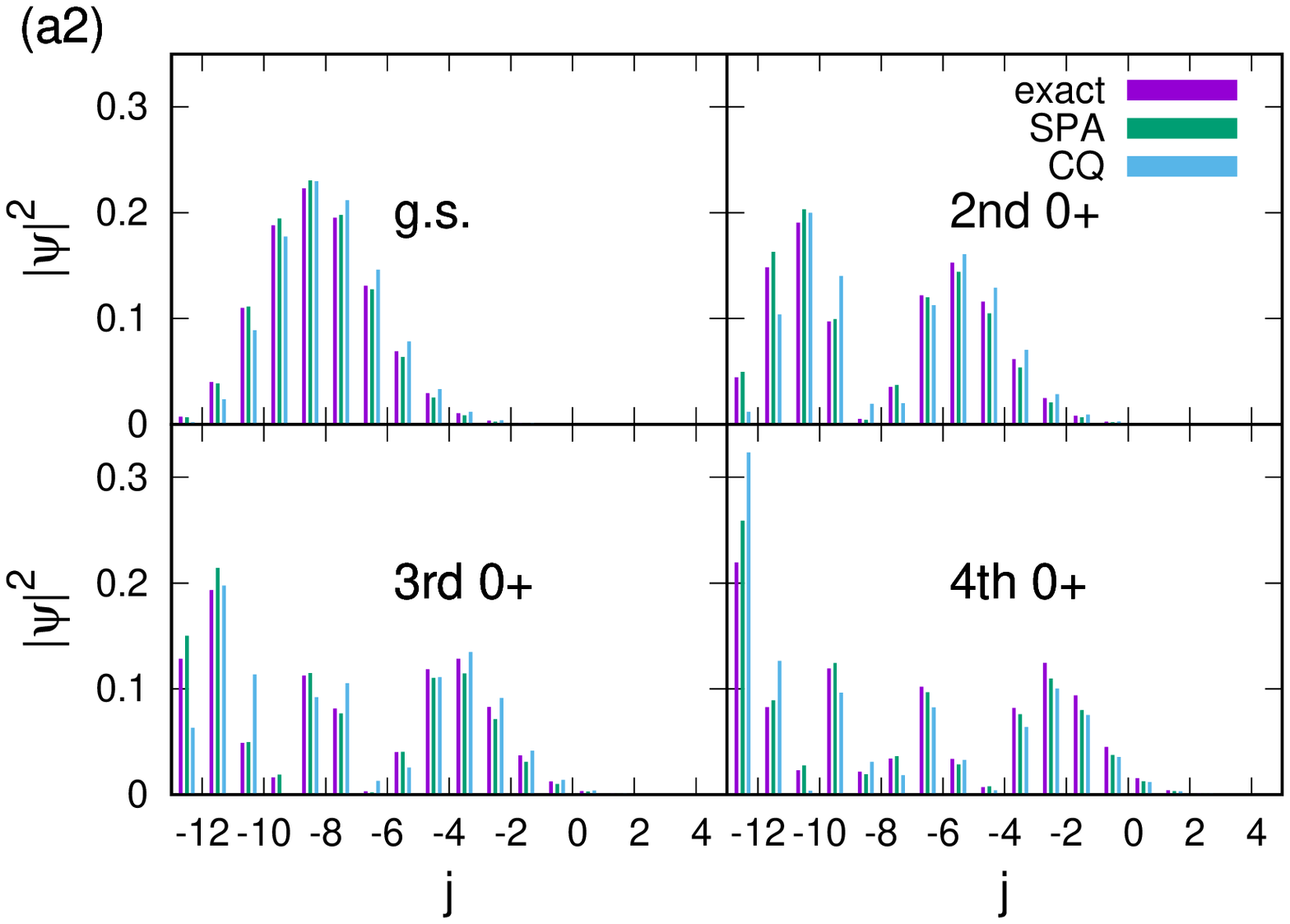}
  \includegraphics[height=0.45\textwidth,angle=-90]{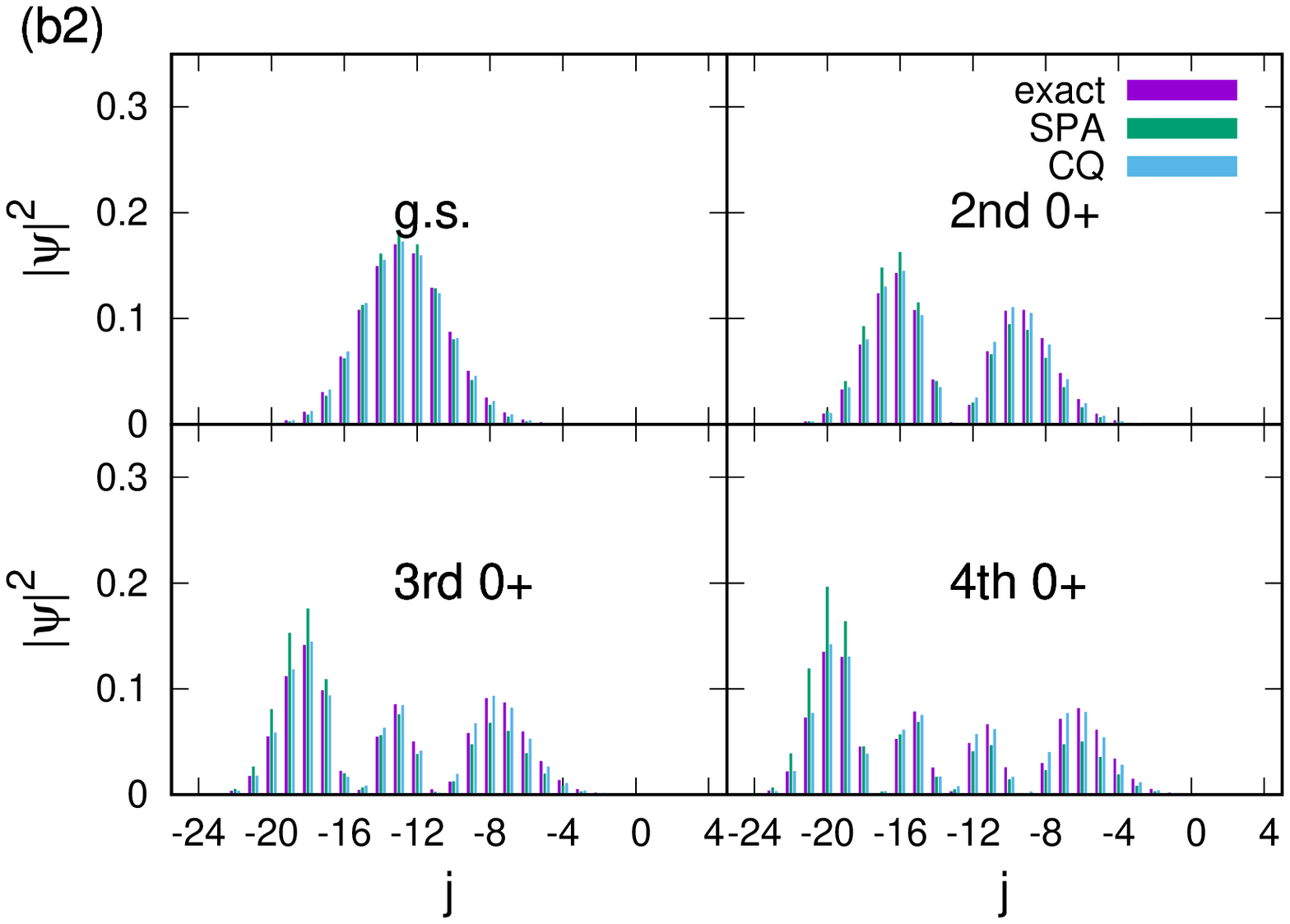}
 \end{center}
 \end{minipage}
 \caption{
Occupation probability in excited $0^+$ states
as a function of $j$ for $\Omega=50$ systems
with (a) $N=50$ and (b) $N=100$.
The upper and lower panels display the results for $x=0.5$ and $x=2$,
respectively. 
The three vertical bars at each $j$ from the left to the right represent
the squared components of the wave functions
from exact, SPA, and CQ calculations, respectively.
The left end of the horizontal axis at $j=j_{\rm min}$
corresponds to a component with $(n_1,n_2)=(N,0)$.
The next at $j=j_{\rm min}+1$ corresponds to the one with
$(n_1,n_2)=(N-2,2)$, and so on.
}
 \label{fig:N50_occ}
\end{figure*}

The obtained wave functions in the SPA and the CQ can be decomposed in the
$2n$-particle-$2n$-hole components in Fig.~\ref{fig:N50_occ}.
In the SPA, it is done as Eq. (\ref{semi_wave_func1}) and
the normalized squared coefficients $|C_m^{(E_k,J)}|^2$ are plotted
in Fig.~\ref{fig:N50_occ}.
For the CQ, $|c_{k,j}^{(N)}|^2$ in Eq. (\ref{CQ_eigenstate}) are shown.
Here, $m$ and $j$ are related to each other, $2j=J-2m$.
They show excellent agreement with the exact results, not only for the
ground state but for excited states.
We find the SPA is even more precise than the CQ.

Next, let us discuss the transition matrix elements.
In this paper, we discuss only $k=0$ (ground state) and $k=1$
(1st excited $\nu=0$ state).
The FD calculation is based on the time evolution of the expectation 
value $S^+(t)$ with fixed $(J,\Phi)$ in Eq. (\ref{Fourier_decomposition}).
For $(N,k)\rightarrow (N+2,k')$ transitions,
we basically adopt the trajectories for the initial state, namely,
the one with $J=N/2$ satisfying the $k$-th EBK quantization condition.
The $k\rightarrow k$ ($\Delta k = 0$)  transitions
correspond to the intraband transitions of the
pair-rotational band, when the state is deformed in the
gauge space (pair deformation).
For the ground-state band ($k=0$),
this is nothing but the expectation value at the BCS wave function,
with the constant value of $S^+$.
Since the constant $S^+$ provides only $\Delta k=0$ intraband transitions,
for the interband transition of $(N,k=0)\rightarrow (N+2,k=1)$
transitions, the trajectory satisfying the EBK condition of $k=1$ is used to
perform the Fourier decomposition (\ref{Fourier_decomposition})
of $\omega=2\pi/T$.

\begin{figure*}[t]
 \begin{minipage}{0.3\hsize}
 \begin{center}
  \includegraphics[width=70mm,angle=-90]{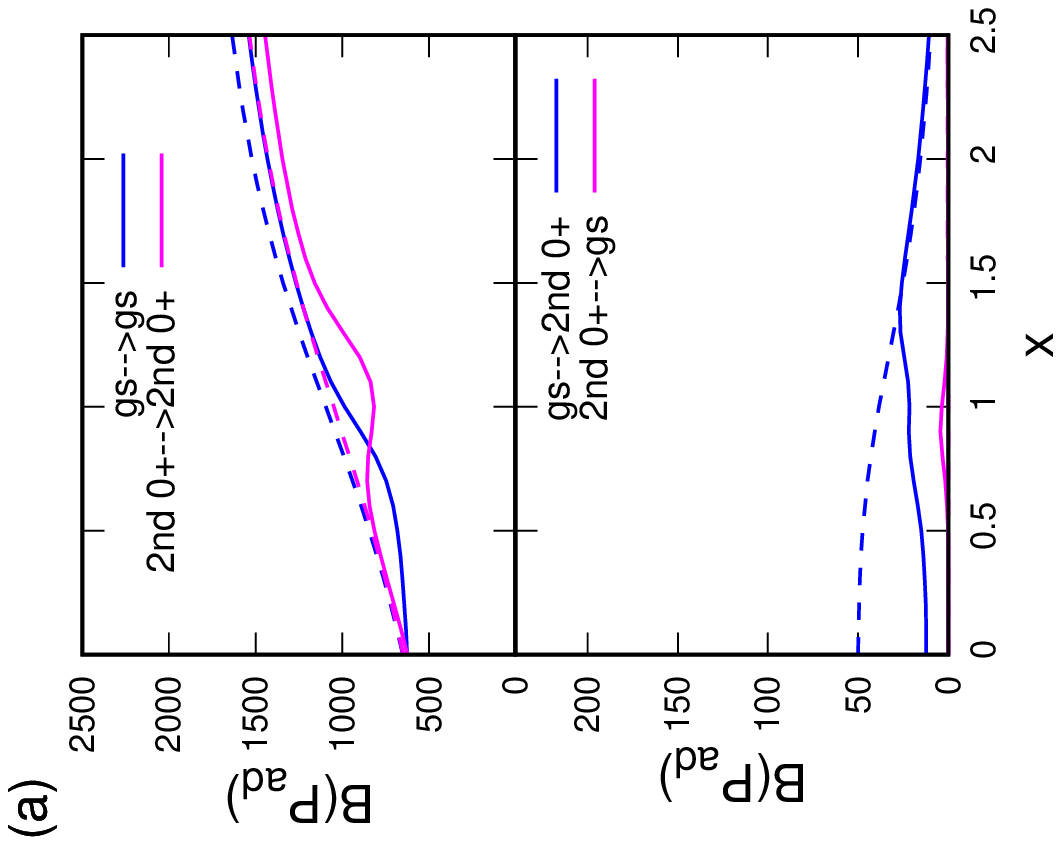}
 \end{center}
 \captionsetup{labelformat=empty,labelsep=none}
 \end{minipage}
 \begin{minipage}{0.3\hsize}
 \begin{center}
  \includegraphics[width=70mm,angle=-90]{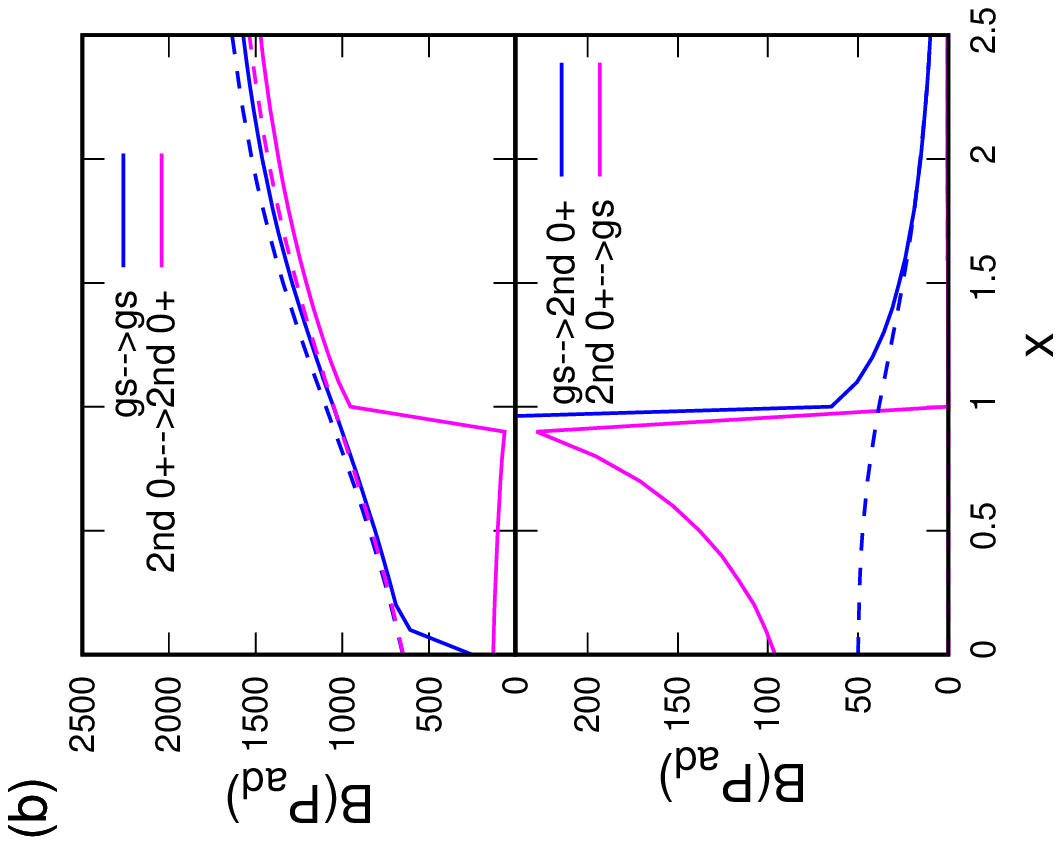}
 \end{center}
 \captionsetup{labelformat=empty,labelsep=none}
 \end{minipage}
 \begin{minipage}{0.3\hsize}
 \begin{center}
  \includegraphics[width=70mm,angle=-90]{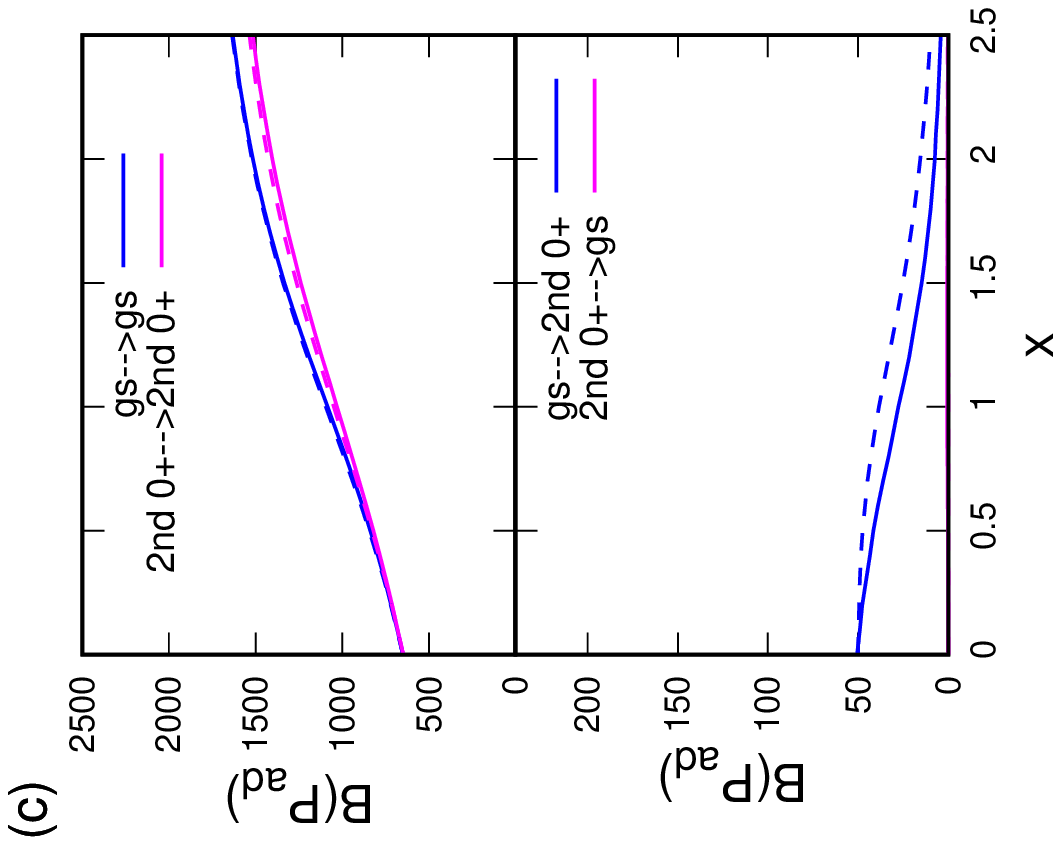}
 \end{center}
 \captionsetup{labelformat=empty,labelsep=none}
 \end{minipage}
 \caption{The strength of pair additional transition
$B(P_{\rm ad};k\rightarrow k')$ for $\Omega=50$ systems
from $N=48$ to 50.
Left panels: results of the CQ method; Middle panels: FD; Right panels: SPA.
Dashed lines represent exact calculation.
Upper panels show the intraband transitions of
$\ket{0_1^+}\to\ket{0_1^+}$ and $\ket{0_2^+}\to\ket{0_2^+}$,
while lower panels show the interband transition of
$\ket{0_1^+}\to\ket{0_2^+}$ and $\ket{0_2^+}\to\ket{0_1^+}$.
}
 \label{fig:N50Pad}
\end{figure*}



\begin{figure*}[t]
 \begin{minipage}{0.3\hsize}
 \begin{center}
  \includegraphics[width=70mm,angle=-90]{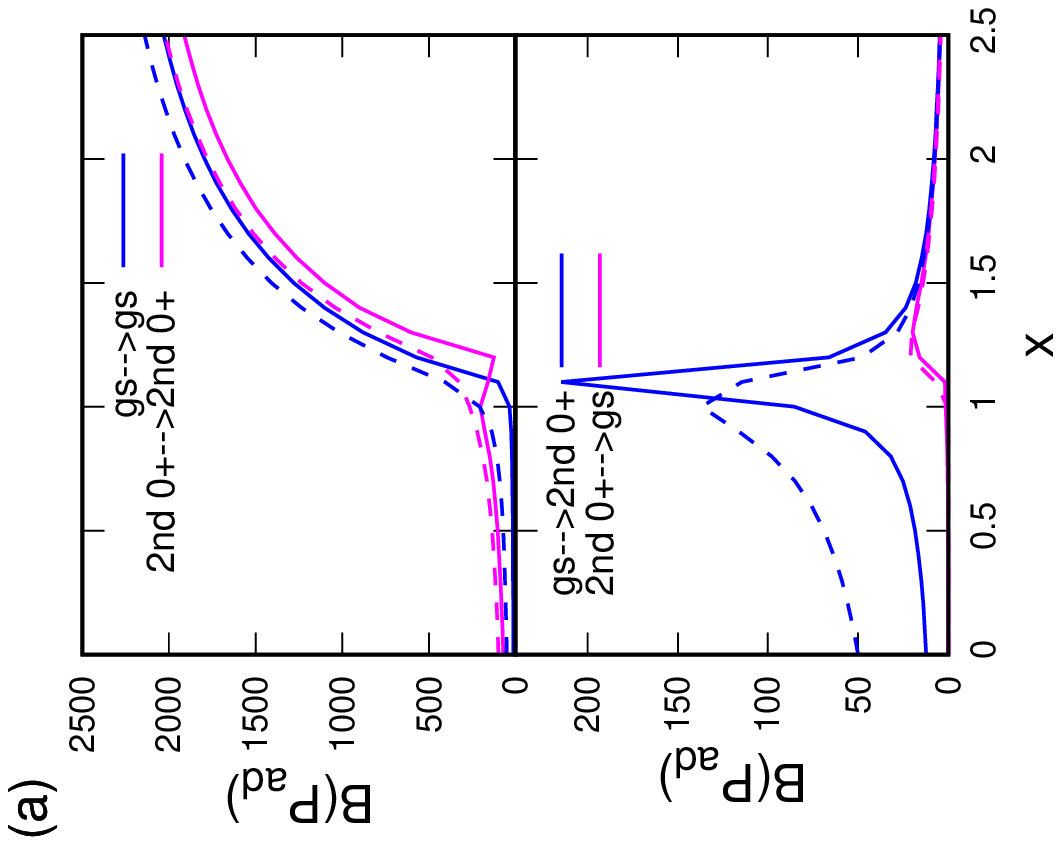}
 \end{center}
 \captionsetup{labelformat=empty,labelsep=none}
 \end{minipage}
 \begin{minipage}{0.3\hsize}
 \begin{center}
  \includegraphics[width=70mm,angle=-90]{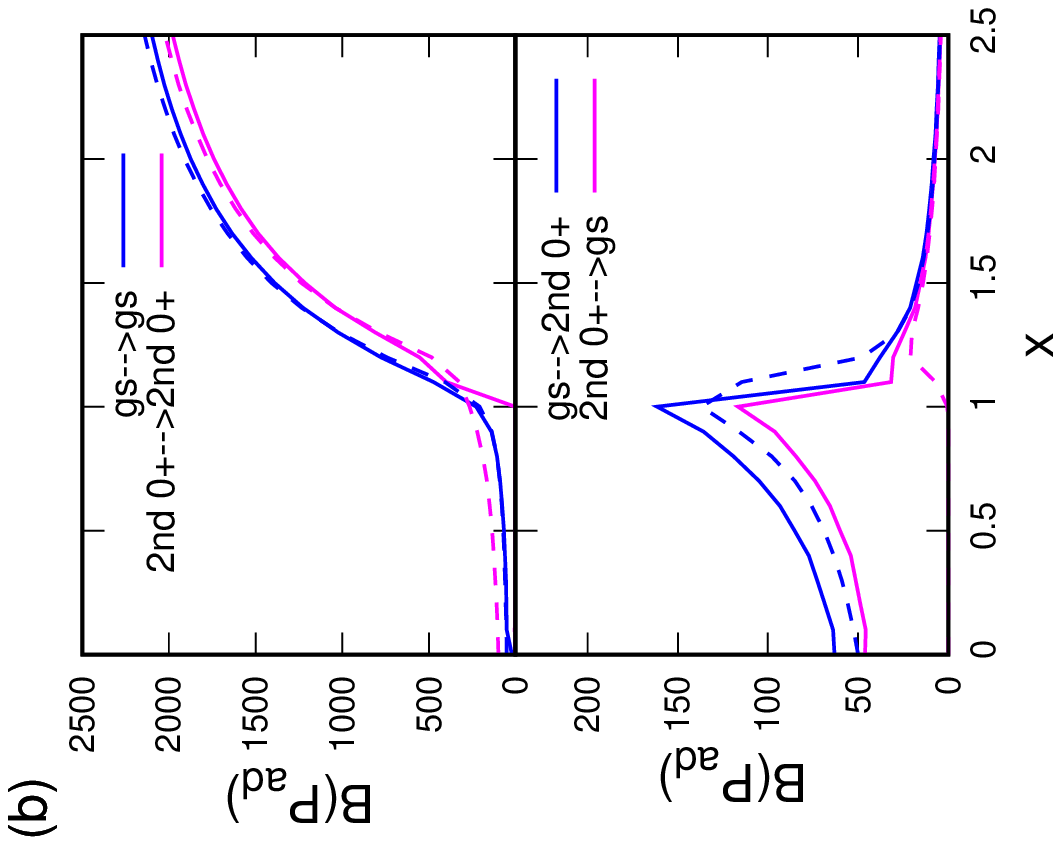}
 \end{center}
 \captionsetup{labelformat=empty,labelsep=none}
 \end{minipage}
 \begin{minipage}{0.3\hsize}
 \begin{center}
  \includegraphics[width=70mm,angle=-90]{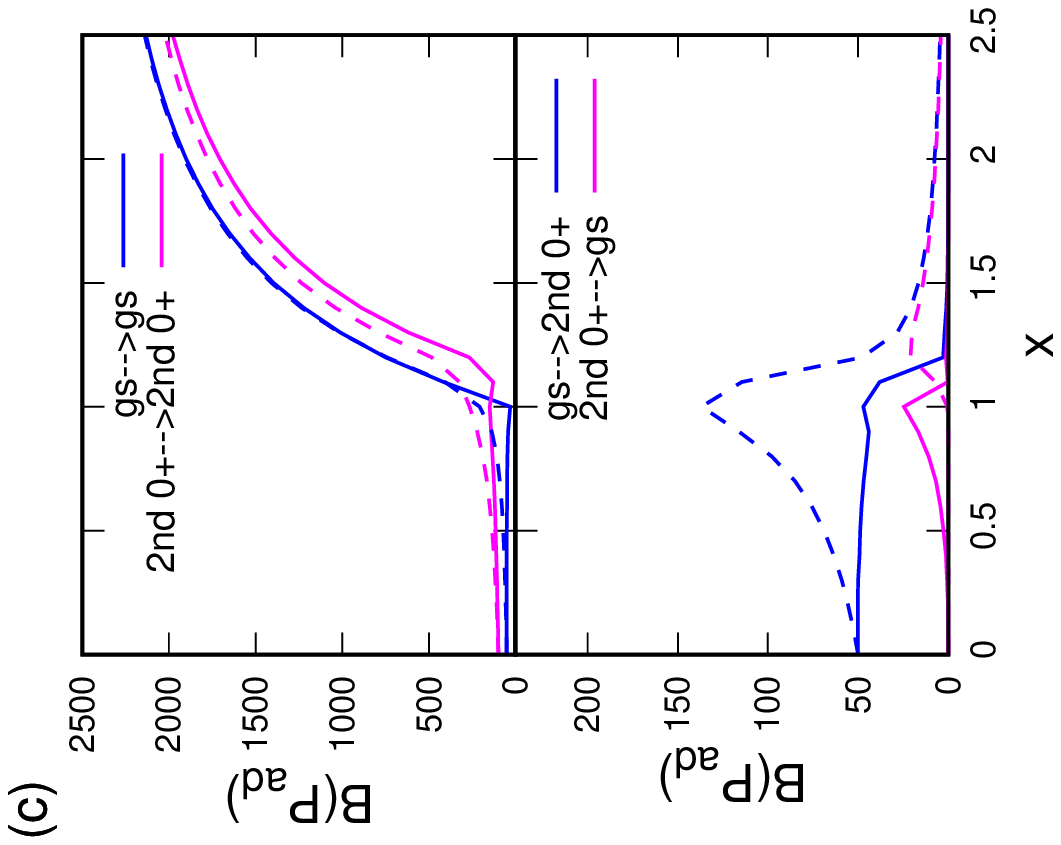}
 \end{center}
 \captionsetup{labelformat=empty,labelsep=none}
 \end{minipage}
 \caption{
The same as Fig.~\ref{fig:N50Pad} but for $N=98\rightarrow 100$.
}
 \label{fig:N100Pad}
\end{figure*}

The calculated pair-addition strengths $B(P_{\rm ad})$ are shown in
Fig.~\ref{fig:N50Pad} for $N=48\rightarrow 50$,
and in Fig.~\ref{fig:N100Pad} for $N=98\rightarrow 100$.
Near the closed-shell configuration ($N=98\rightarrow 100$),
the pair-addition strengths for the intraband transitions ($\Delta k=0$)
drastically increase around $x=1$.
This reflects a character change from the pair vibration ($x\lesssim 1$)
to the pair rotation ($x\gtrsim 1$).
The $B(P_{\rm ad}; k\rightarrow k)$ in the pair-rotational transitions
are about 20 times larger than those in the vibrational transitions.
The interband $B(P_{\rm ad}; 0\rightarrow 1)$ are similar to
the $B(P_{\rm ad}; 0\rightarrow 0)$ in the vibrational region
($x\lesssim 1$), because they both change the number of pair-phonon quanta
by one unit.
In contrast, $B(P_{\rm ad}; 1\rightarrow 0)$, which change the phonon
quanta by three, are almost zero.
In the pair-rotational region ($x \gg 1$),
$B(P_{\rm ad}; 1\rightarrow 0)$ and $B(P_{\rm ad}; 0\rightarrow 1)$
are roughly identical. This is because both $B(P_{\rm ad}; 1\rightarrow 0)$ 
and $B(P_{\rm ad}; 0\rightarrow 1)$ correspond to one-phonon excitation
in ``deformed'' cases ($x \gg 1$).

In the mid-shell region ($N=48\rightarrow 50$),
the intraband $B(P_{\rm ad}; k\rightarrow k)$ are smoothly increase as
$x$ increases.
Their values are larger than the interband strengths by about one (two) order
of magnitude at $x\sim 0$ ($x\sim 2.5$),
indicating the pair-rotational character.
The interband $B(P_{\rm ad}; 0\rightarrow 1)$ show a gradual decrease
as a function of $x$, while
$B(P_{\rm ad}; 1\rightarrow 0)$ are negligibly small,
even at $x\gg 1$.
This presents a prominent difference from the closed-shell case.

All the features of the pair-transfer strengths
are nicely reproduced in the SPA method,
for both the closed- and mid-shell configurations.
The CQ method qualitatively agrees with the exact calculation.
For instance, the order-of-magnitude difference between intraband
and interband transitions.
However, the precision of the CQ method is not so good,
especially around $x=1$.
The FD method properly describes the main features
in the superfluid phase, while it fails for the normal phase
($x\lesssim 1$). 
In the mid-shell configuration, the ground state is always
in the superfluid phase at $x>0$,
while the $k=1$ excited state corresponds to
the open (closed) trajectory at $0<x\lesssim 1$ ($x\gtrsim 1$).
For the open trajectory, the FD produces wrong values.
However, somewhat surprisingly,
the SPA, which uses these open trajectories for the construction
of wave functions,
reproduces main features of the exact results.

\subsection{Small-$\Omega$ cases}

Next, we discuss systems with smaller degeneracy $\Omega=8$.
Again, we study systems near the closed-shell and the mid-shell configurations.

\begin{figure}[t]
 \begin{center}
  \includegraphics[height=0.46\textwidth,angle=-90]{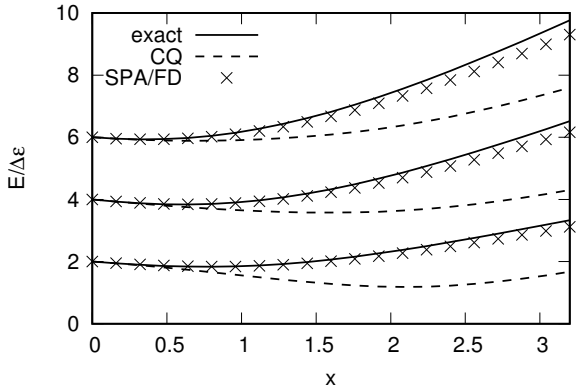}
 \end{center}
 \caption{Excitation energies of $\ket{0_2^+}$, $\ket{0_3^+}$ and
$\ket{0_4^+}$ for $\Omega=N=8$ systems as functions of $x$.
}
 \label{fig:N8energy}
\end{figure}

\begin{figure}[htbp]
 \begin{minipage}{1\hsize}
 \begin{center}
   \includegraphics[height=0.9\textwidth,angle=-90]{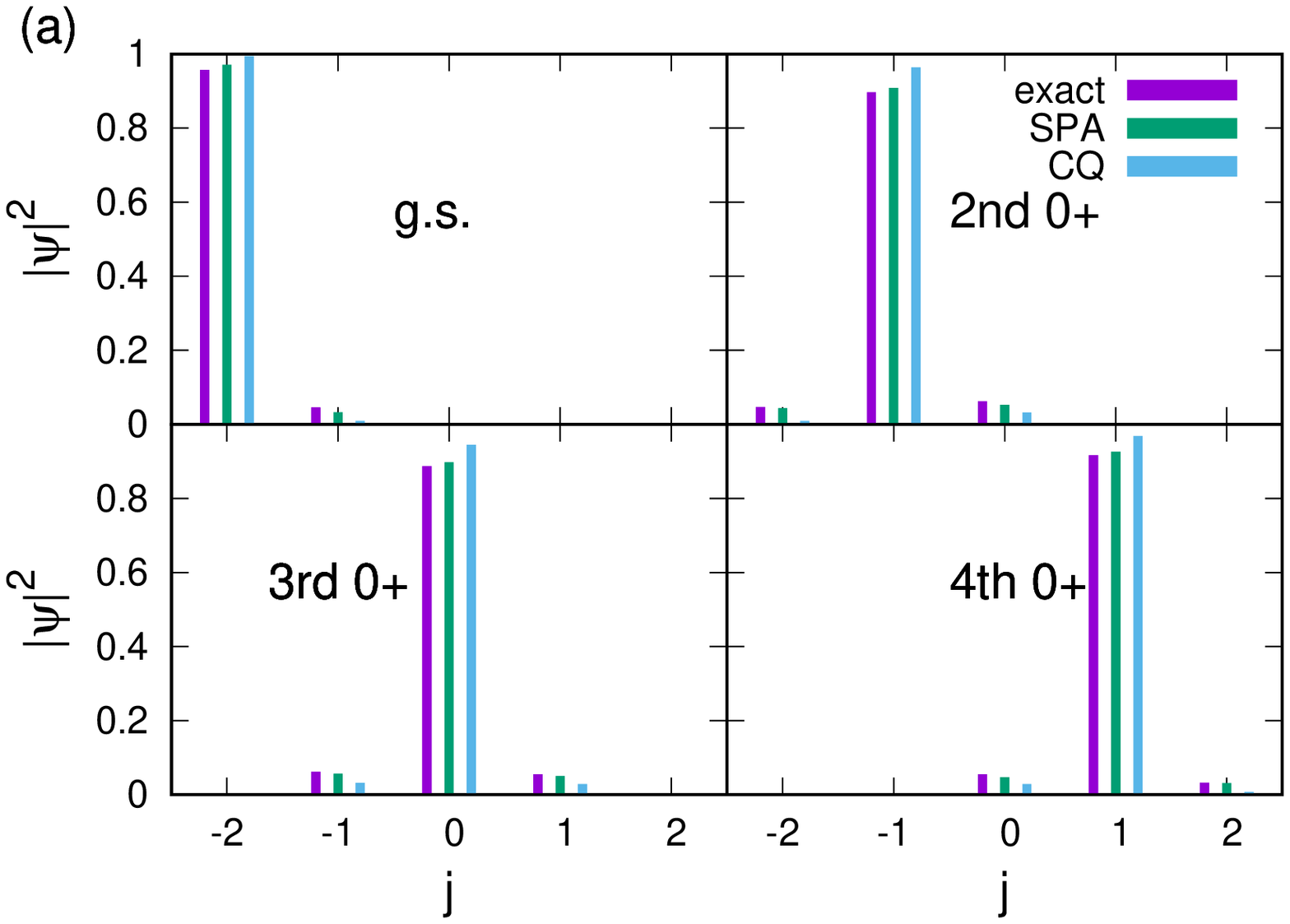}
   \includegraphics[height=0.9\textwidth,angle=-90]{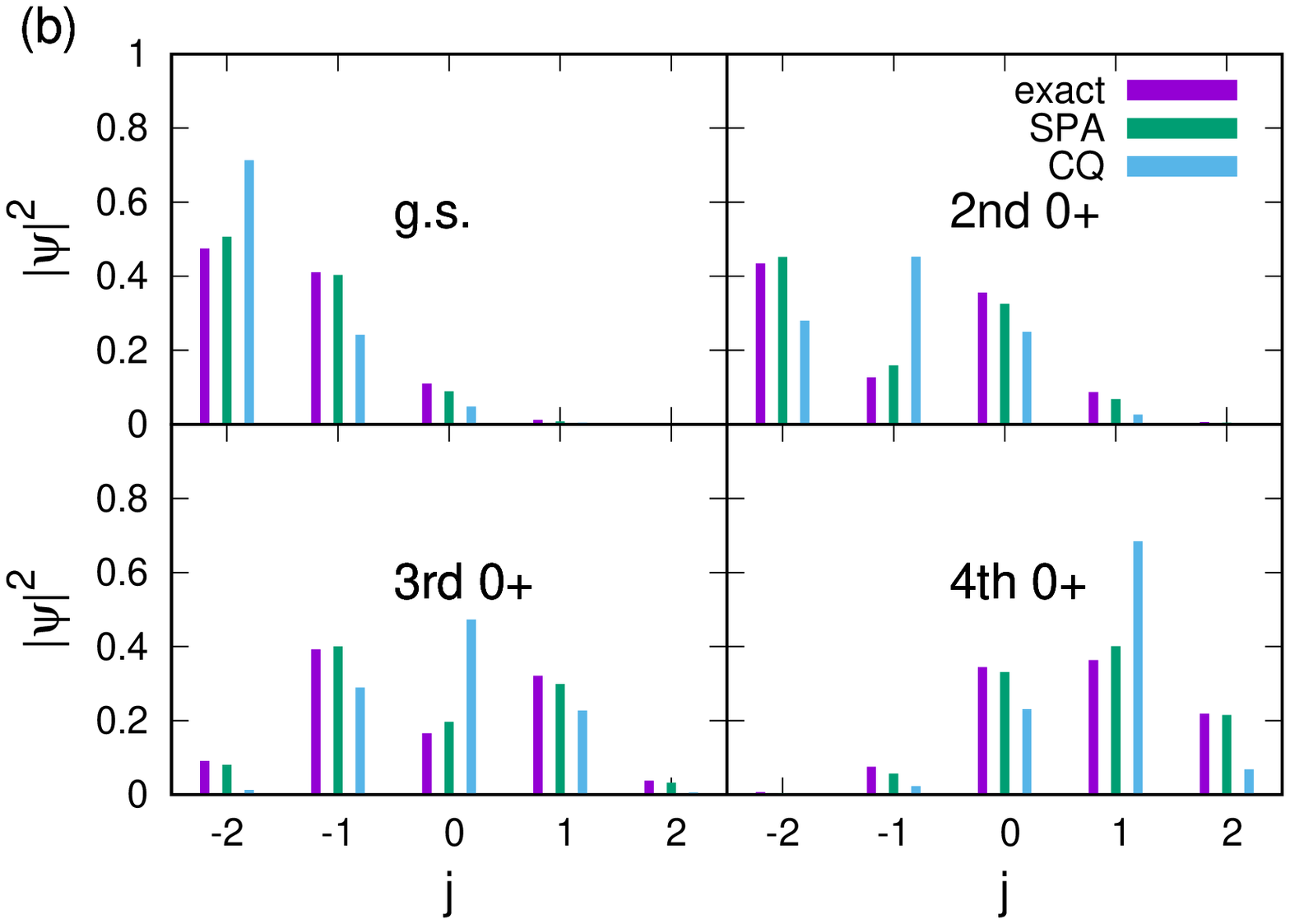}
 \end{center}
 \end{minipage}
 \caption{
Occupation probability in excited $0^+$ states
as a function of $j$ for $\Omega=N=8$ systems.
The panels (a) and (b) display the results for $x=0.5$ and $x=2$,
respectively. 
See also the caption of Fig.~\ref{fig:N50_occ}.
}
 \label{fig:N8_occ}
\end{figure}

\subsubsection{Mid-shell configuration}

The calculated excitation energies are shown in Fig.~\ref{fig:N8energy}
for the $N=8$ case.
The SPA/FD reproduces the exact calculation in the entire region of $x$,
not only for the lowest but also for higher excited states.
The CQ reproduces the exact result in a weak pairing region ($x\lesssim 1$),
while it underestimates the excitation energies at $x\gtrsim 1$.
Analogous to the case of $\Omega=50$,
this is mainly due to effect of the zero-point energy $\Delta E$.
The ground-state energy in the CQ calculation 
is bound at $x\gtrsim 1$.
However, because of the weak collectivity with $N=8$,
the first excited state stays unbound even at the maximum $x$ in 
Fig.~\ref{fig:N8energy}.
Therefore, the energy shift $\Delta E>0$ is larger in the ground state,
which makes the excitation energy smaller.

The wave functions are plotted in Fig.~\ref{fig:N8_occ}. 
At the weak pairing case of $x=0.5$,
both the SPA and the CQ reproduce the exact result.
At $x=2$, the squared coefficients of the ground state has an
asymmetric shape peaked at the
lowest $j$, which suggests that the state is not deeply bound in the
potential.
It is in contrast to the symmetric shape in Fig.~\ref{fig:N50_occ}.
The wave functions obtained by the CQ method has noticeable deviation
from the exact results.
On the other hand, the SPA wave functions are almost identical to the
exact ones.

\begin{figure*}[t]
 \begin{minipage}{0.3\hsize}
 \begin{center}
  \includegraphics[width=70mm,angle=-90]{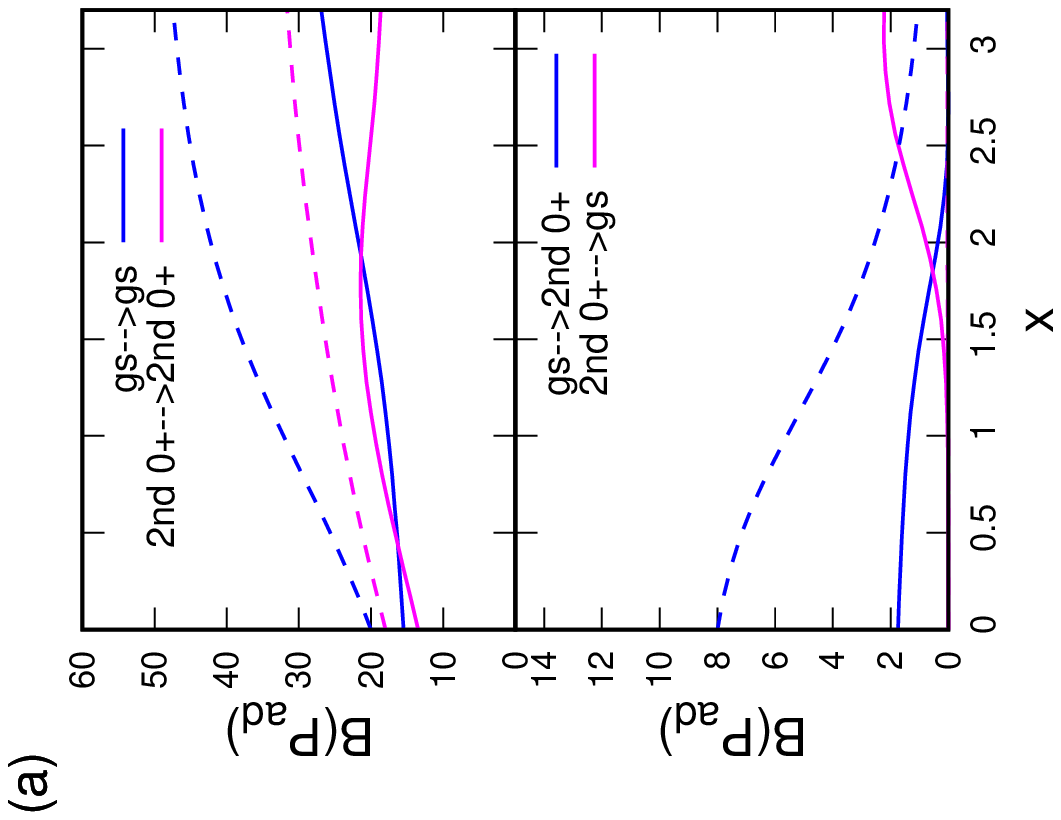}
 \end{center}
 \captionsetup{labelformat=empty,labelsep=none}
 \end{minipage}
 \begin{minipage}{0.3\hsize}
 \begin{center}
  \includegraphics[width=70mm,angle=-90]{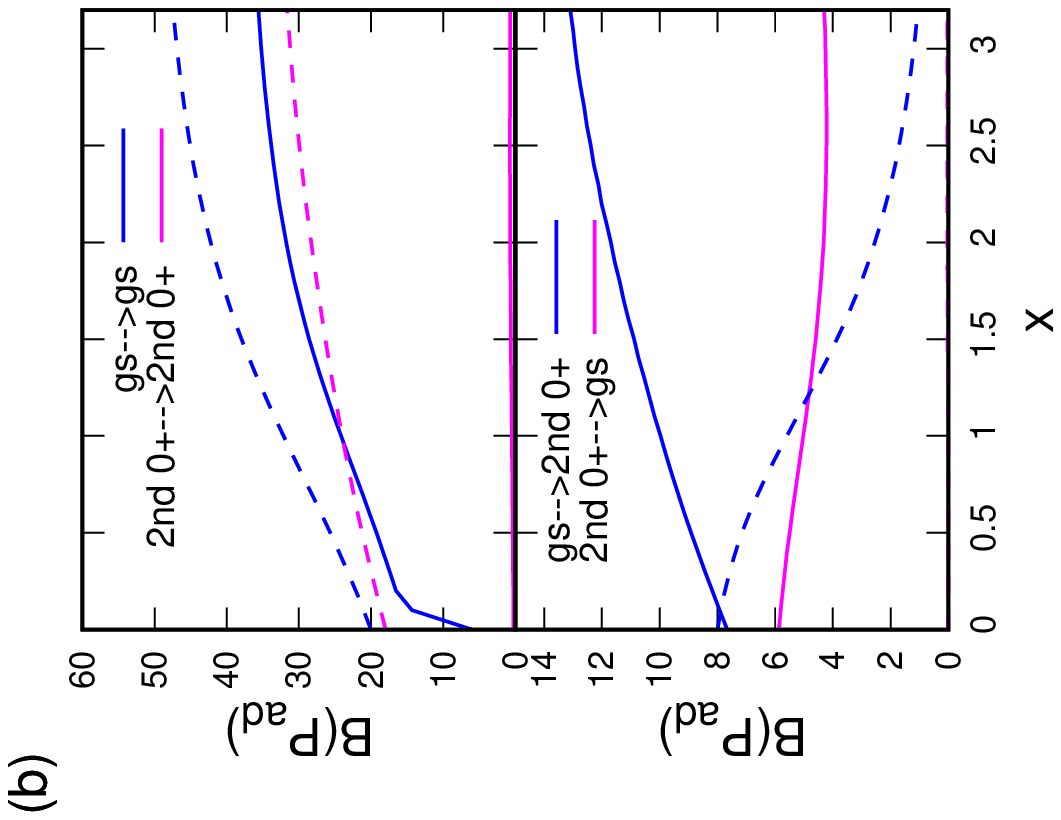}
 \end{center}
 \captionsetup{labelformat=empty,labelsep=none}
 \end{minipage}
 \begin{minipage}{0.3\hsize}
 \begin{center}
  \includegraphics[width=70mm,angle=-90]{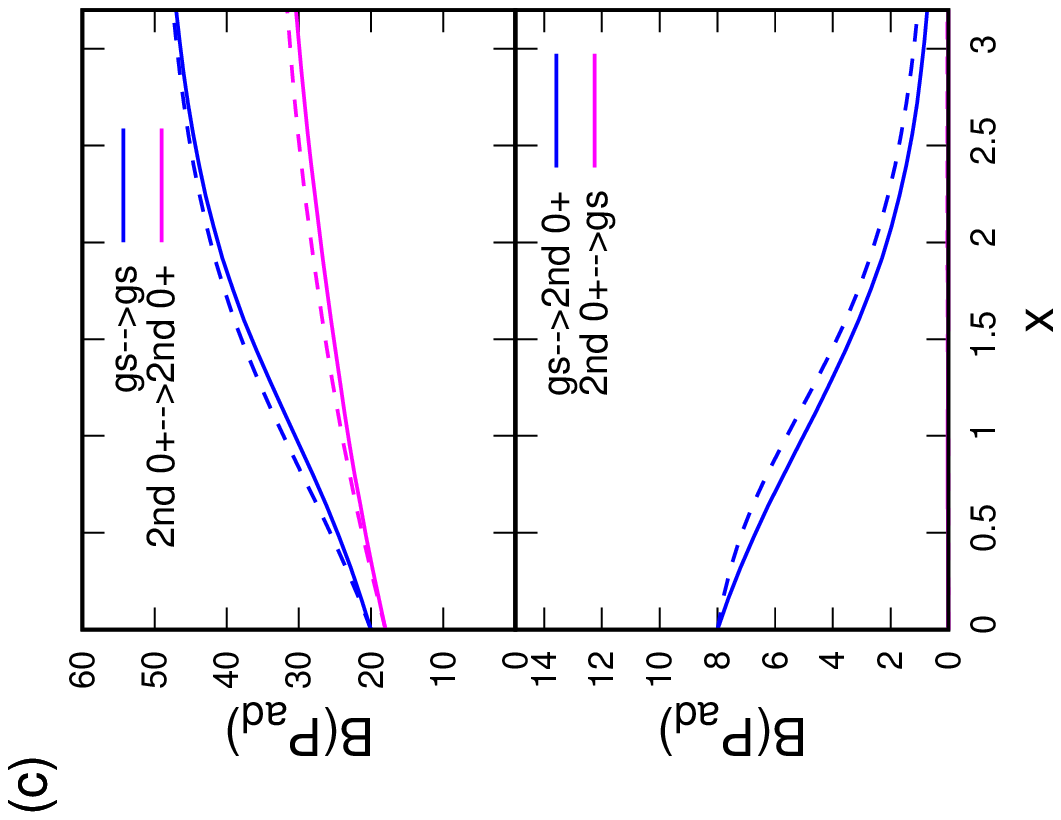}
 \end{center}
 \captionsetup{labelformat=empty,labelsep=none}
 \end{minipage}
	\caption{The same as Fig.~\ref{fig:N50Pad} but for $N=6\rightarrow 8$
	with $\Omega=8$.
}
 \label{fig:N8Pad}
\end{figure*}

The pair-addition transition strengths from $N=6$ to $N=8$ are
shown in Fig.~\ref{fig:N8Pad}.
The intraband $k\rightarrow k$ transitions increase and 
the interband $k=0\rightarrow 1$ transitions decrease as
functions of $x$.
Their relative difference becomes
more than one order of magnitude at $x\gtrsim 2$.
Thus, even at relatively small $\Omega$ and $N$,
the intraband transitions in the pair rotation is qualitatively different
from the interband transitions.

We find the excellent agreement between the SPA and the exact calculations.
The first excited state corresponds to the open trajectory which
turns out to almost perfectly reproduce the exact wave function.
In contrast, this open trajectory produces results far from the exact
one in the FD method.
It produces almost vanishing the intraband
$B(P_{\rm ad}; 1\rightarrow 1)$.
The $B(P_{\rm ad}; 0\rightarrow 0)$ shows a qualitative agreement for
its behavior, but is significantly underestimated.
The CQ method also underestimates the intraband transitions.

For the mid-shell configurations, the SPA is dominantly superior to the
CQ and the FD methods.


\begin{figure}[htb]
 \begin{center}
  \includegraphics[height=0.46\textwidth,,angle=-90]{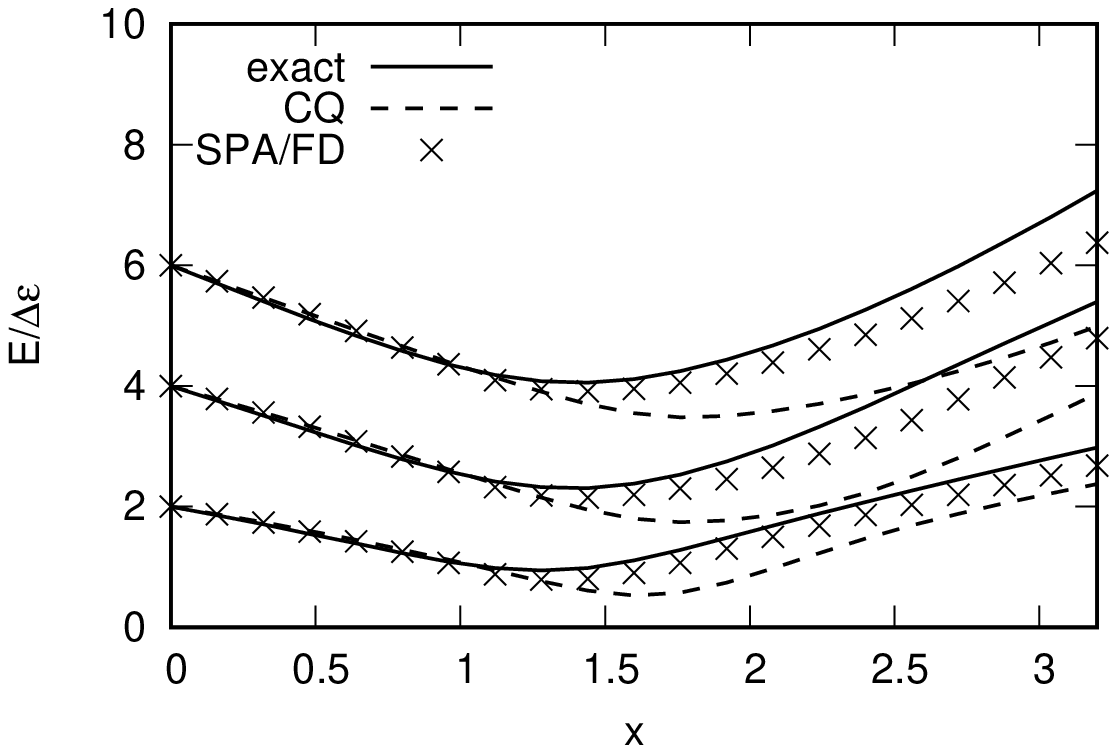}
 \end{center}
	\caption{The same as Fig.~\ref{fig:N8energy} but for
	$N=2\Omega=16$.
}
 \label{fig:N16energy}
\end{figure}

\begin{figure}[thb]
 \begin{minipage}{1\hsize}
 \begin{center}
   \includegraphics[height=0.9\textwidth,angle=-90]{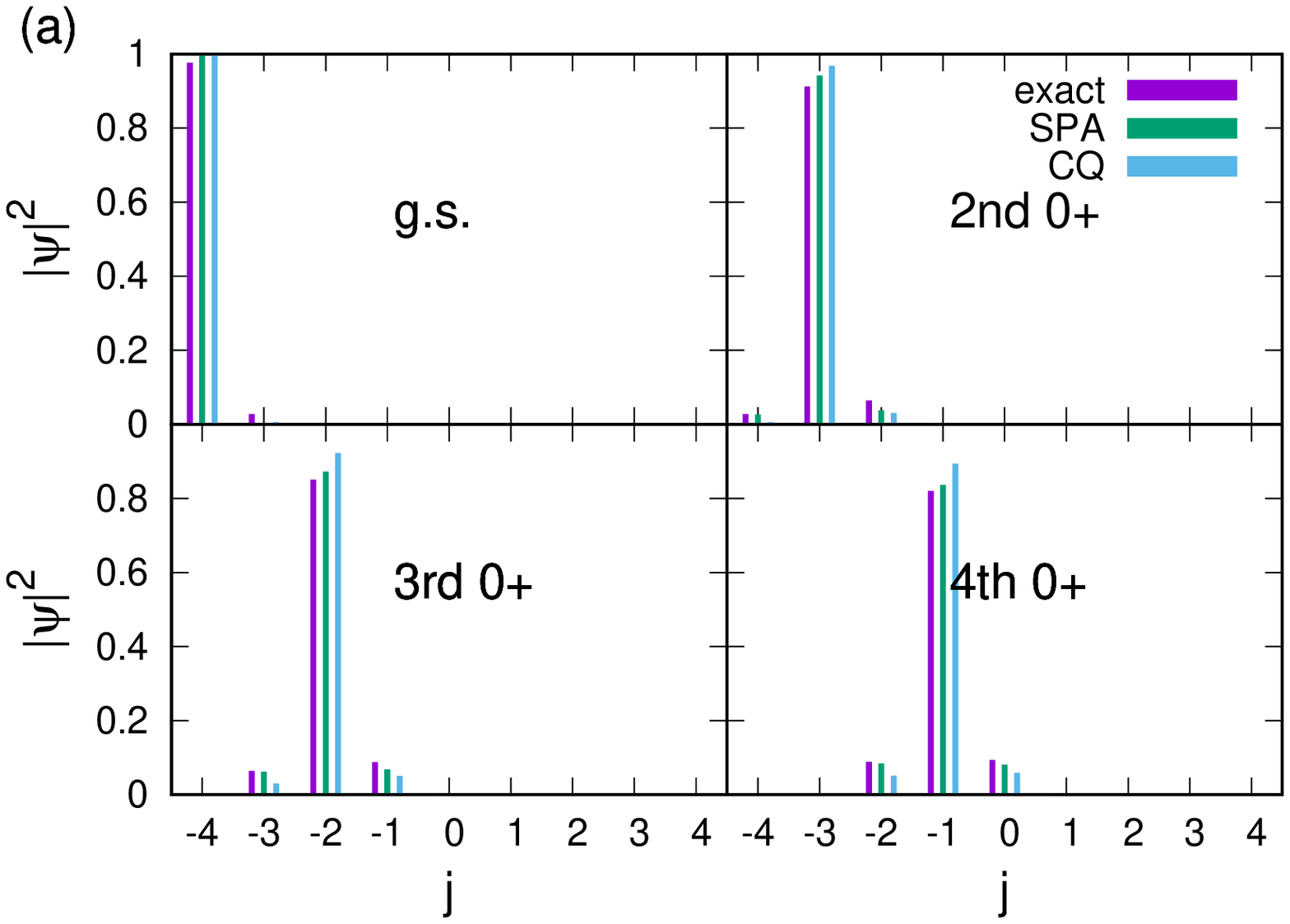}
   \includegraphics[height=0.9\textwidth,angle=-90]{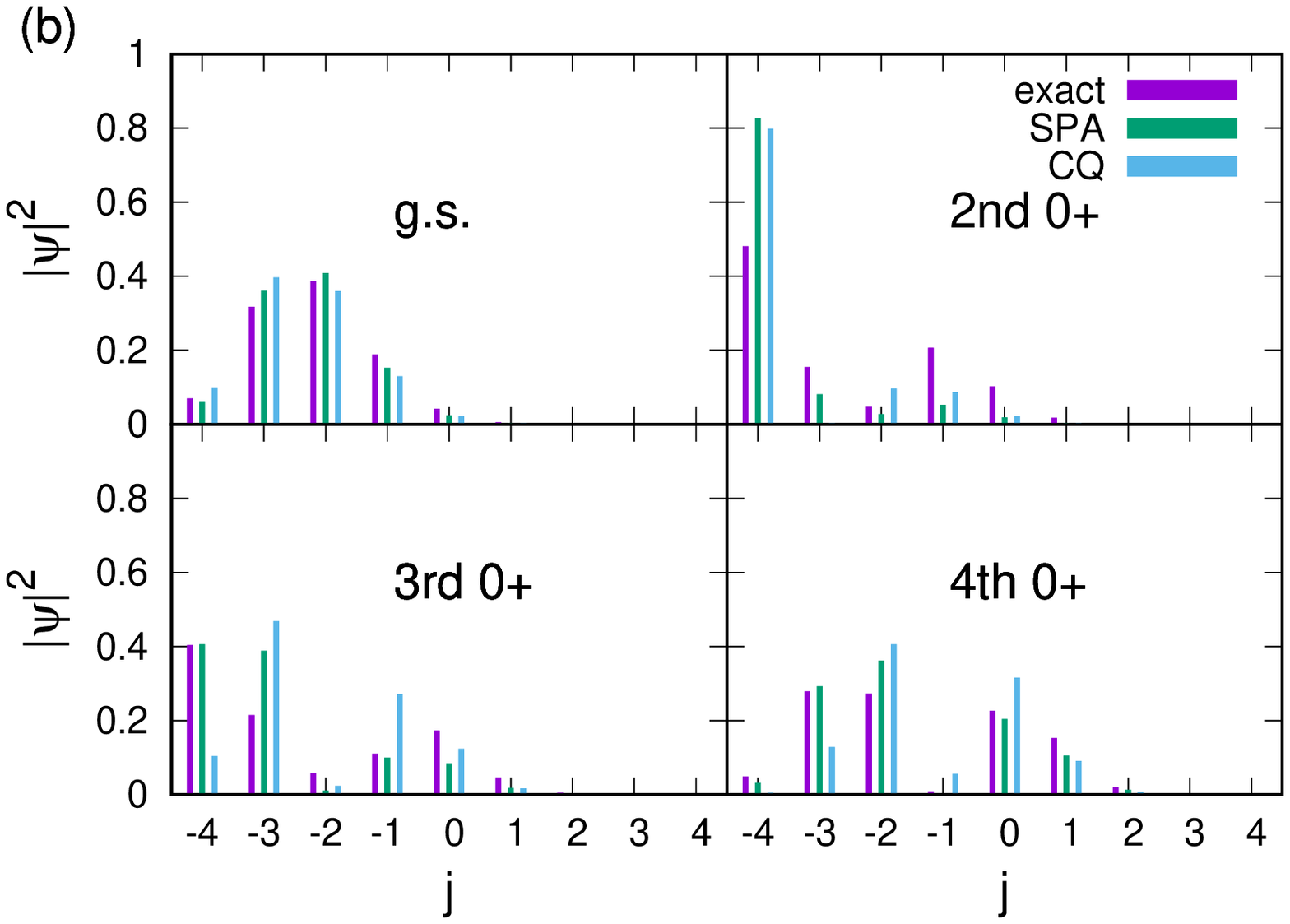}
 \end{center}
 \end{minipage}
	\caption{The same as Fig.~\ref{fig:N8_occ} but for
	$N=2\Omega=16$.
}
 \label{fig:N16_occ}
\end{figure}

\subsubsection{Closed-shell configuration}

In the closed shell with $N=16$,
the minimum-energy trajectory changes at $x=1$ from
$j=-4$ (normal phase) to
the BCS minimum $j>-4$ and $\phi=0$ (superfluid phase).
At the transitional point ($x=1$), the harmonic approximation is
known to collapse, namely to produce zero excitation energy.
In Fig.~\ref{fig:N16energy},
this collapsing is avoided in all the calculations (SPA/FD and CQ).
The behaviors of the lowest excitation agree with the exact calculations,
while the CQ method substantially underestimates those for higher states.
This is again due to the difference in the zero-point energy in
the ground and the excited states.
In the CQ calculation, the first excited state is bound at $x\gtrsim 2$,
but the second excited state is unbound for $x\lesssim 3.2$.

Near the transition point from the open to closed trajectories,
the wave functions calculated with the SPA and CQ methods somewhat differ
from the exact ones.
In Fig.~\ref{fig:N16_occ}, the wave functions at $x=0.5$ and 2 are presented.
They agree with exact calculation at $x=0.5$.
In contrast, we find some deviations for the first excited state ($k=1$)
at $x=2$.
This is because the $k=1$ trajectory corresponding to the first excited state
changes its character from open to closed at $x\approx 1.8$.
Therefore, the first excited wave function is difficult to reproduce in
the SPA, although the wave functions for
the ground and higher excited states show reasonable agreement.
The similar disagreement is observed for the ground state near $x=1$.

\begin{figure*}[t]
 \begin{minipage}{0.3\hsize}
 \begin{center}
 \includegraphics[width=70mm,angle=-90]{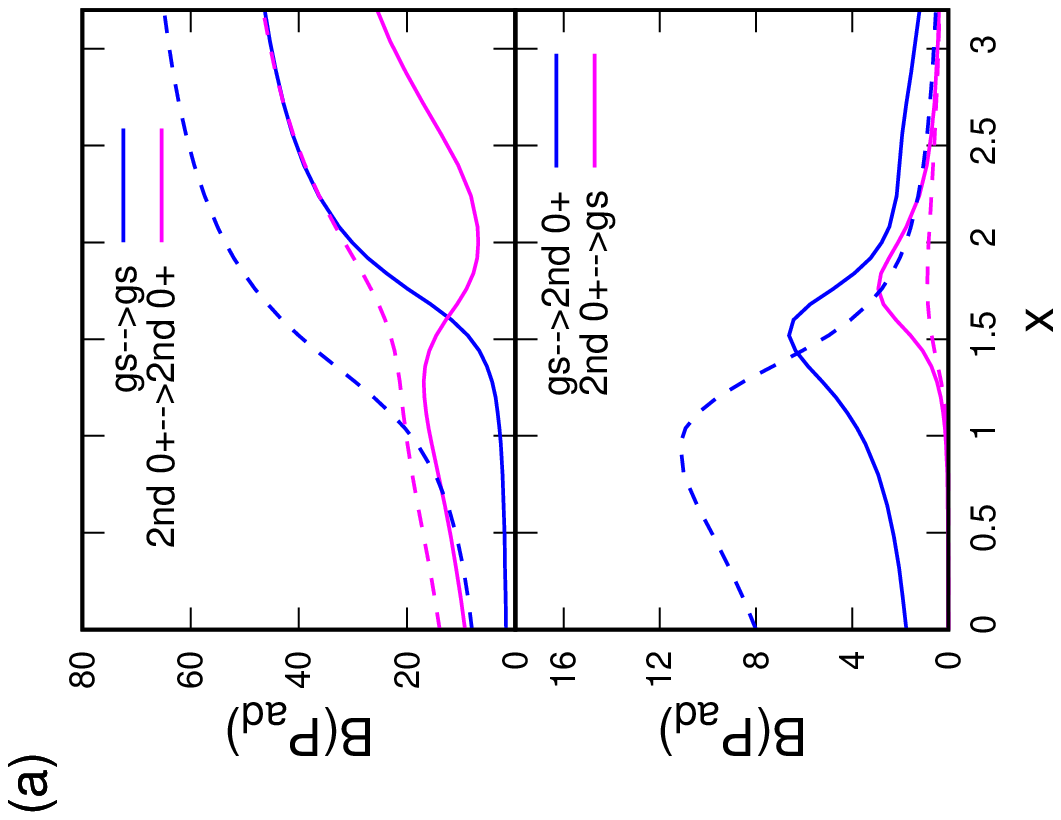}
 \end{center}
 \captionsetup{labelformat=empty,labelsep=none}
 \end{minipage}
 \begin{minipage}{0.3\hsize}
 \begin{center}
 \includegraphics[width=70mm,angle=-90]{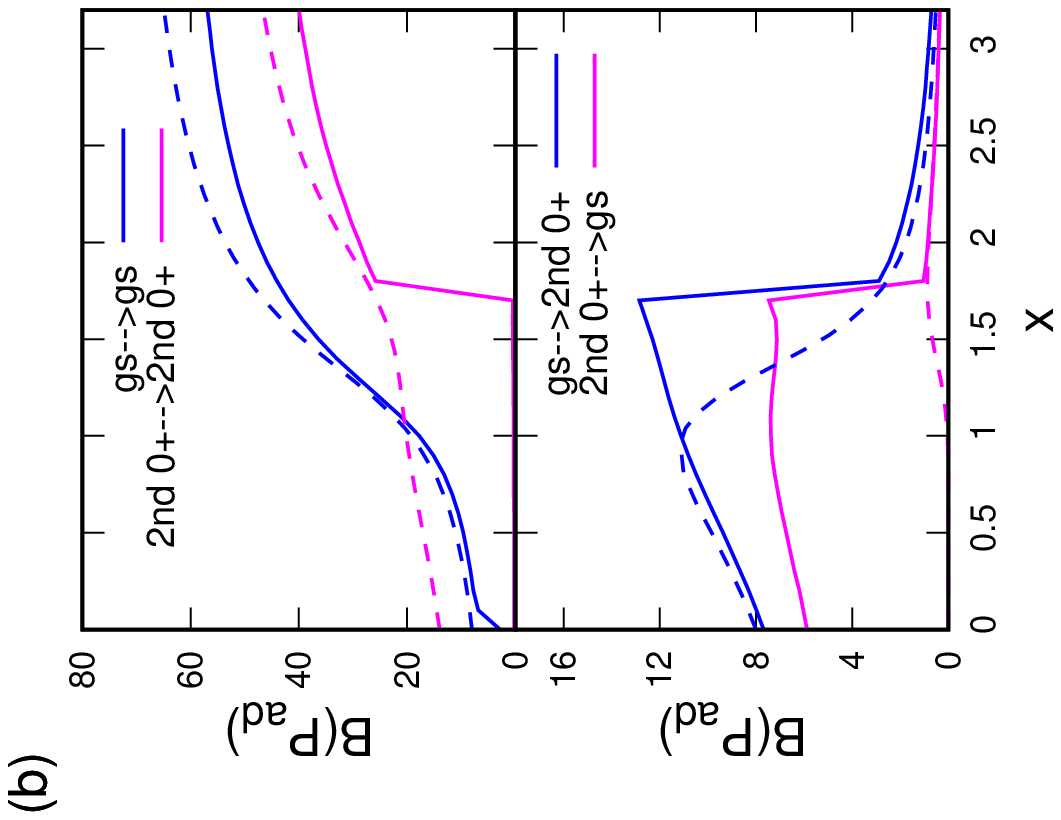}
 \end{center}
 \captionsetup{labelformat=empty,labelsep=none}
 \end{minipage}
 \begin{minipage}{0.3\hsize}
 \begin{center}
 \includegraphics[width=70mm,angle=-90]{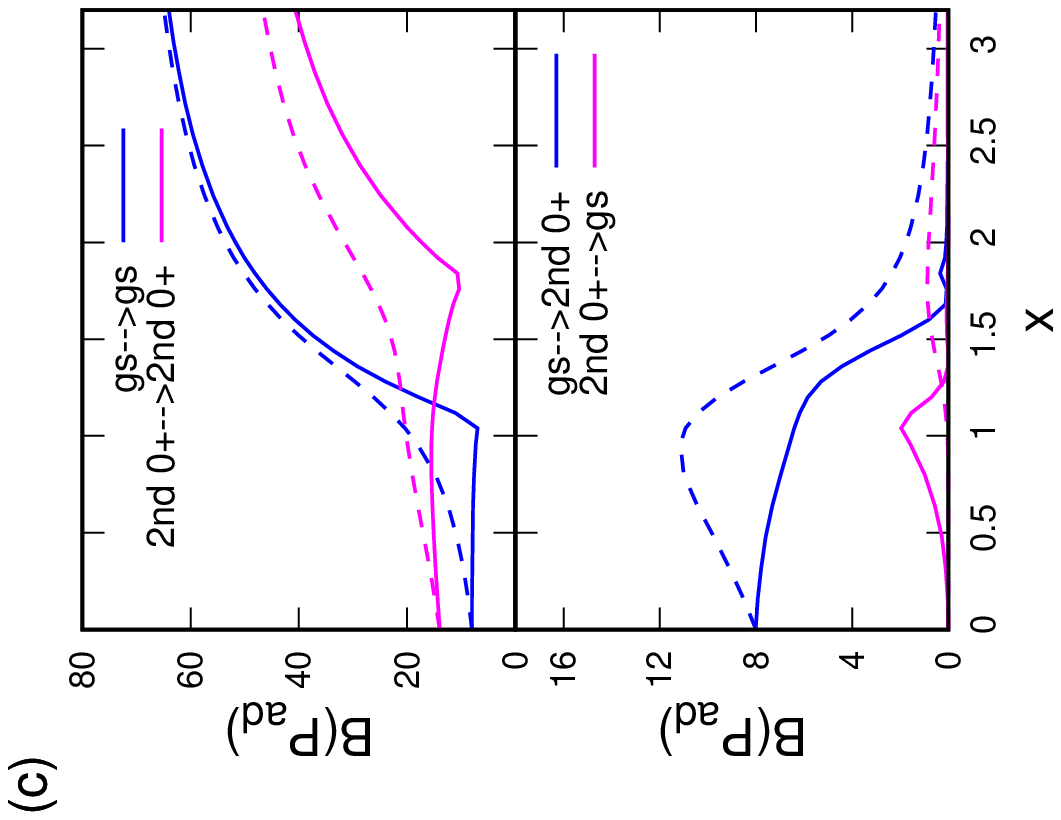}
 \end{center}
 \captionsetup{labelformat=empty,labelsep=none}
 \end{minipage}
	\caption{The same as Fig.~\ref{fig:N50Pad} but for $N=14\rightarrow 16$
	with $\Omega=8$.
}
 \label{fig:N16Pad}
\end{figure*}

Singular behaviors near the transition points can be also observed in
the pair-addition transition strengths ($N=14\rightarrow 16$) shown
in Fig.~\ref{fig:N16Pad}. 
At $x=1$, the intraband $B(P_{\rm ad}; 0\rightarrow 0)$ shows a kink
in the SPA,
and $B(P_{\rm ad}; 1\rightarrow 1)$ shows another kink at $x\approx 1.8$.
These exactly correspond to the transition points from open to closed
trajectories.
Nevertheless, the overall behaviors are well reproduced and
the values at the weak and strong pairing limit are reasonably reproduced
in the SPA.
The CQ calculation also shows smoothed kink-like behaviors near the
transition points.
However, it underestimates the intraband $B(P_{\rm ad}; k\rightarrow k)$.
The FD method does not have a kink for $B(P_{\rm ad}; 0\rightarrow 0)$,
because $S^+(t)$ is calculated for an $N=14$ system.
Both intraband and interband transitions in the FD calculations
reasonably agree with the exact results at $x\gtrsim 1.8$.
The $k=1$ state is not properly reproduced at $x\lesssim 1.8$
with the open trajectory.

For the closed-shell configurations,
the SPA and the FD methods provide reasonable description for the
pair-transfer transition strengths.



\subsection{Collective model treatment}

The collective model has been proposed
and utilized for the nuclear pairing dynamics
\cite{BBPK70,delta1,delta3}.
For those studies, the pairing gap parameter (or equivalent quantities)
is assumed to be the collective coordinates.
This is analogous to the five-dimensional (5D) collective (Bohr) model,
in which the collective coordinates are assumed to be the
quadrupole deformation parameters, $\alpha_{2\mu}$.
The 5D collective model has been 
extensively applied to analysis on numerous experimental data.
On the contrary, there have been very few applications of
the pairing collective model
in comparison with experimental data.
In this section, we examine the validity of
the collective treatment of the pairing.


Although the global gauge angle $\Phi$ is arbitrary,
the deformation parameter $\alpha\equiv \braket{ \hat{S}^- }$
in the gauge space is
usually taken as a real value ($\Phi=0$).
The energy minimization with a fixed value of real $\alpha$ always
leads to $\phi=0$.
\begin{align}
\alpha(j,J) = \braket{Z_0|\hat{S}^-|Z_0} 
	= \frac{\Omega}{2} \left(\sqrt{1-q_1^2} +\sqrt{1-q_2^2} \right).
 \label{Delta_BCS}
\end{align}
The parameter $\alpha$ is equivalent to the pairing gap $\Delta$,
since the relation, $\Delta=G\alpha$,
guarantees one-to-one correspondence between $\alpha$ and $\Delta$.
In Sec.~\ref{sec:canonical}, we treat $\phi$ as a collective coordinate
and $j$ as its conjugate momentum.
The collective model treatment is based on the opposite choice,
$j$ as a coordinate and $\phi$ as a momentum.

\begin{figure}[tb]
 \begin{center}
(a) \includegraphics[height=0.44\textwidth,angle=-90]{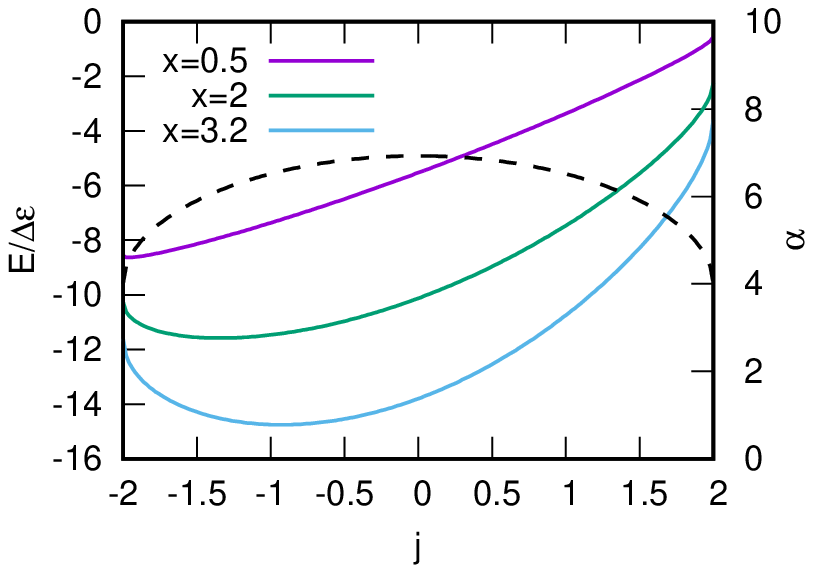}
\\
(b) \includegraphics[height=0.44\textwidth,angle=-90]{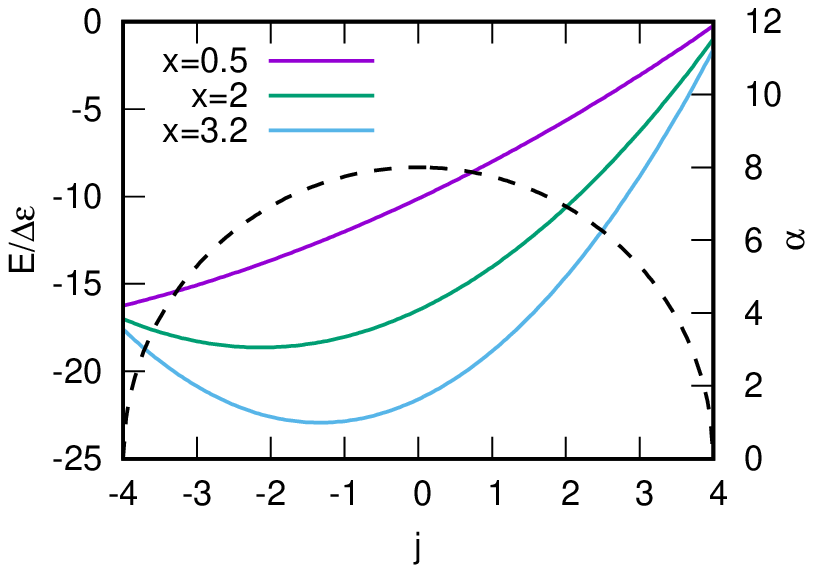}
 \end{center}
 \caption{Energy surfaces as functions of $j$,
 Eq. (\ref{TDHFB_Hamiltonian_3}) with $\phi=0$,
 for $x=0.5$, $2$, and $3.2$.
 Dashed line is the pairing deformation parameter $\alpha$
 of Eq.~(\ref{Delta_BCS}) as a function of $j$.
 (a) mid-shell ($\Omega=N=8$) and (b) closed-shell ($2\Omega=N=16$).
}
 \label{fig:p_Delta}
\end{figure}


The problem is that
there is no one-to-one correspondence between $j$ and $\alpha$.
The relation between $j$ and $\alpha$ 
are shown by dashed lines in Fig.~\ref{fig:p_Delta}
for $\Omega=8$ mid-shell (a) and closed-shell (b) configurations.
The deformation parameter $\alpha$ is largest at $j=0$ (equal filling in both levels),
and smallest at the end points of $j$.
The constrained minimization with respect to $\alpha$ cannot 
produce the states corresponding to $j>0$.
Apparently, we cannot map the entire region of $j$ to $\alpha$.

The collective model treatment
requires the collective wave functions to be well localized
in the $j<0$ region.
The potential energy, $E(j)=\mathcal{H}(\phi=0,j;J=N/2)$ of
Eq. (\ref{TDHFB_Hamiltonian_3}), is also shown in
Fig.~\ref{fig:p_Delta}.
The restriction becomes more serious for the stronger pairing cases.
For instance, the potential with $x=3.2$ in Fig.~\ref{fig:p_Delta}(a)
has only about 1 MeV depth at the minimum point, relative to the value
at the boundary point ($j=0$) corresponding to
the maximum value of $\alpha$ ($\Delta$).

\begin{figure}[tb]
 \begin{center}
(a) \includegraphics[height=0.4\textwidth,angle=-90]{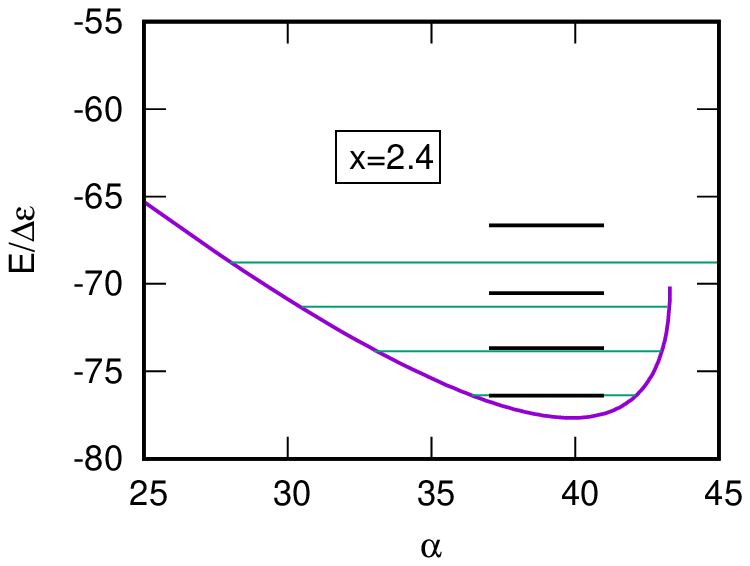}
\\
(b) \includegraphics[height=0.4\textwidth,angle=-90]{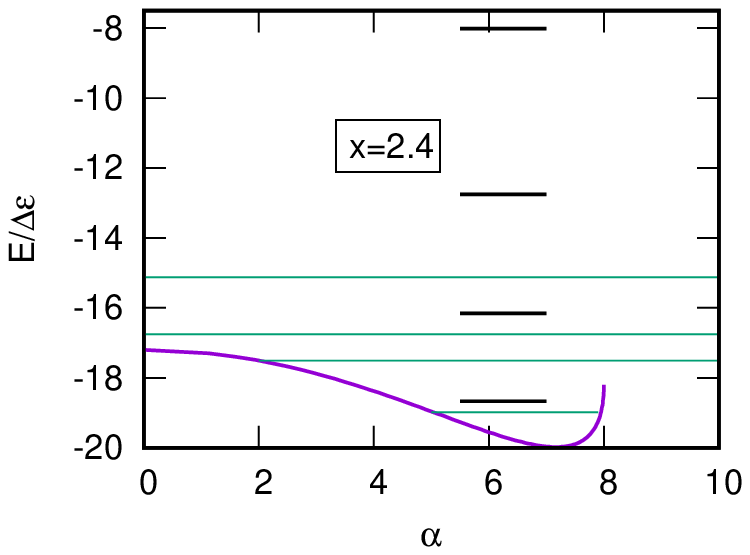}
 \end{center}
\caption{Potential energy surface as functions of $\alpha$ with $x=2.4$;
(a) $\Omega=N=50$,
(b) $2\Omega=N=16$.
Horizontal lines 
indicate energy spectra. Black lines are obtained
from the potential energy surface, and green lines are 
obtained with the CQ method
(Sec.~\ref{sec:canonical}).
}
 \label{fig:Delta_E}
\end{figure}

To simulate the result of the collective model, we expand the Hamiltonian
(\ref{TDHFB_Hamiltonian_3}) up to the second order in $\phi$,
then, quantize it by $\hat{\phi}=i\partial/\partial j$ with
the ordering given by Pauli's prescription.
The range of the coordinate $j$ is restricted to $j_{\rm min}\leq j \leq 0$
with the vanishing boundary condition $\psi(j_{\rm min})=\psi(0)=0$.
Figure~\ref{fig:Delta_E} shows two examples of the relationship 
between excitation energies and the potential energy surface.
In the panel (a), we show the case of large $\Omega$
($\Omega=50$, $N=50$, and $x=2.4$),
in which
the excited $0^+$ states are bound up to second excitation.
From the collective model, the energies of the ground and the first excited states
are well described, while the deviation becomes larger for
higher excited states.
For very large degeneracy, the pocket of energy surface is deep, 
hence the low-lying excited states may be described by $\alpha$. 
However, in the small-$\Omega$ case ($\Omega=8$, $N=16$, and $x=2.4$)
of the panel (b), no excited states are bound by the potential
as a function of $\alpha$.
None of the excited $0^+$ states
are properly described in the collective model.
This shallow potential is a consequence of the improper
choice of the collective coordinate $\alpha$
which represents only the $j<0$ region.
Therefore, the collective model treatment assuming
$\alpha$ ($\Delta$) as the collective coordinate
is not applicable to small-$\Omega$ and strong-pairing cases.

\newpage

\section{Conclusion}
\label{sec:conclusion}

The different methods of the requantization of the TDHFB dynamics
was studied for the two-level pairing model;
the stationary-phase approximation (SPA) of the path integral,
the canonical quantization (CQ), and the Fourier decomposition (FD)
of the time-dependent observables.
In this model, since the global gauge angle $\Phi$ is a cyclic variable,
the TDHFB dynamics can be described by the integrable classical dynamics.
After the pair-rotation variables $(\Phi,J)$ are separated,
the remaining degrees of freedom, $(\phi,j)$, describe
the pair-vibrational motion.

In systems with large degeneracy $\Omega$ and number of particles $N$,
all the quantization methods reasonably reproduce the results of the
exact calculation for excitation spectra.
It is more difficult to reproduce the two-particle transfer matrix elements.
Nevertheless, for the large $\Omega$ and $N$,
we obtain qualitative agreement with the exact results.
These are ideal cases,
but realistic situations may have smaller $\Omega$ and $N$
in the valence space.

In systems with relatively small $\Omega$ and $N$,
the agreement is less quantitative for the CQ and the FD,
especially for the two-particle transfer matrix elements.
In contrast, the SPA keeps its accuracy in the entire range of
pairing strengths.
One of the reasons of its success is due to the inclusion of the off-diagonal 
parts of the pair transfer operator,
by the explicit construction of the microscopic wave functions.
The CQ and FD calculate the pair transfer matrix elements using only
the diagonal part (expectation value) of the operator $\hat{S}^\pm$,
based on Eqs. (\ref{Sp_mean_value}) and (\ref{Sm_mean_value}).
This is a good approximation when the collectivity is so large that the
diagonal parts dominate.
However, the pairing collectivity may be too weak to justify this
treatment.


We also investigated the conventional treatment of
the collective model which assumes that the collective coordinate is
the paring gap parameter.
As we mentioned before, the present two-level pairing Hamiltonian has
only one pair of collective variables $(\phi,j)$, in addition to the
pair-rotational variables $(\Phi,J)$.
Even for such a simple system, we find that it is difficult to justify
the use of $\Delta$ as the collective coordinate,
especially for relatively small-$\Omega$ cases.
Basically, there is no one-to-one correspondence between $\Delta$ and $j$.
The collective wave functions are not necessarily bound in the region
where the variable $\Delta$ can represent.


Among the different requantization methods,
the SPA is the most accurate tool for description of
the pairing large amplitude collective motion in realistic nuclear systems.
The weak point of this approach is that it is applicable only to the
integrable TDHFB system.
In order to solve this problem, we plan to first extract the
integrable collective submanifold in the many-dimensional TDHFB phase space.
For this purpose, the adiabatic self-consistent collective coordinate (ASCC)
method \cite{NMMY16} is a promising tool.
The combined study of the ASCC and the SPA for multi-level systems is our
next target under progress.

\begin{acknowledgments}
This work is supported in part
by Interdisciplinary Computational Science Program in CCS,
University of Tsukuba,
and
by JSPS-NSFC Bilateral Program for Joint Research Project
on Nuclear mass and life for unravelling mysteries of r-process.
\end{acknowledgments}

\appendix
\begin{widetext}
\section{Derivation of semiclassical wave function}

We give a derivation of the semiclassical wave function
of Eq. (\ref{semi_wave_func1}).
The explicit form of coherent state is

\begin{align}
  \ket{Z} &= \prod_{l=1}^2(1+|Z_l|^2)^{-\Omega_l/2}e^{Z_lS_l^{+}}\ket{0} \nonumber \\
  &= \prod_{l=1}^2\left(1+\tan^2{\frac{\theta_l}{2}}\right)^{-\Omega_l/2} \sum_k \frac{1}{k!} \sum_{m=0}^k
  \binom{k}{m} \left(
  \tan{\frac{\theta_1}{2}}e^{-i\phi_1}S_1^{+} \right)^m \left(
  \tan{\frac{\theta_2}{2}}e^{-i\phi_2}S_2^{+} \right)^{k-m} \ket{0} \nonumber \\
  &= \prod_{l=1}^2\left(1+\tan^2{\frac{\theta_l}{2}}\right)^{-\Omega_l/2} \sum_k \frac{1}{k!} \sum_{m=0}^k
  \binom{k}{m} \tan^m{\frac{\theta_1}{2}}\tan^{k-m}{\frac{\theta_2}{2}}e^{-ik\Phi}e^{-i(k/2-m)\phi}
  (S_1^{+})^m(S_2^{+})^{k-m} \ket{0}
  \label{ZZ}
\end{align}
Inserting (\ref{ZZ}) into (\ref{semi_wave_func0}) under fixed $N$ and $E_k$,
it becomes
\begin{align}
	\ket{\psi_k^N} \propto& \oint d\Phi \oint dt
	e^{i\mathcal{T}_{N,E_k}(\Phi, t)}
	\ket{\Phi,t}_{N,E_k} \nonumber \\
  \propto& \sum_k \frac{1}{k!} \sum_{m=0}^k \binom{k}{m} \int_0^{2\pi} d\Phi e^{i(N/2-k)\Phi} 
  \int_0^T dt e^{i\int \pi(t') \dot{\phi}(t') dt'-i(k/2-m)\phi} \nonumber \\
  &\times \left\{ \prod_{l=1}^2\left(1+\tan^2{\frac{\theta_l}{2}}\right)^{-\Omega_l/2} \right\}
\tan^m{\frac{\theta_1}{2}}\tan^{k-m}\frac{\theta_2}{2}(S_1^{+})^m(S_2^{+})^{k-m} \ket{0} \nonumber \\
  \propto& \sum_{m=0}^{N/2} \binom{N/2}{m} \int_0^T dt \exp{\left( i\int \pi(t') \dot{\phi}(t') dt'-i(N/4-m)\phi \right)} \nonumber \\
  &\times \left\{ \prod_{l=1}^2\left(1+\tan^2{\frac{\theta_l}{2}}\right)^{-\Omega_l/2} \right\} 
\tan^m{\frac{\theta_1}{2}}\tan^{N/2-m}\frac{\theta_2}{2} (S_1^{+})^m(S_2^{+})^{N/2-m} \ket{0} . \label{semi_wave_func_b}
\end{align}
We find that the integration over $\Phi$ is nothing but the number projection.
In SU(2) quasi-spin representation, the vacuum state is written as
$\ket{0}=\ket{S_1,-S_1;S_2,-S_2}$, which leads to
\begin{align}
  (S_1^{+})^m(S_2^{+})^{N/2-m} \ket{0} 
  = \sqrt{\frac{(2S_1)!m!}{(2S_1-m)!}} \sqrt{\frac{(2S_2)!(N/2-m)!}{[2S_2-(N/2-m)]!}}
  \ket{S_1,-S_1+m;S_2,-S_2+(N/2-m)}.
\end{align}
For convenience, we define coefficients
\begin{align}
  A(q,S,m) &\equiv 
  \frac{\tan^m{\frac{\theta}{2}}}{\left(1+\tan^2{\frac{\theta}{2}}\right)^{-S}}
  \sqrt{\frac{(2S)!m!}{(2S-m)!}}  \nonumber \\
  &= \left(\frac{1-q}{2}\right)^{m/2}\left(\frac{1+q}{2}\right)^{S-m/2} \sqrt{\frac{(2S)!m!}{(2S-m)!}}
\label{tan}
\end{align}
where $q=\cos{\theta}$.
Inserting Eq. (\ref{tan}) into Eq. (\ref{semi_wave_func_b}) with $N/2=J$,
we reach Eq. (\ref{semi_wave_func1}).
\end{widetext} 

\bibliographystyle{apsrev4-1}
\bibliography{ref2}

\end{document}